\documentclass[5p]{elsarticle}

\usepackage{lineno,hyperref}
\usepackage[nolist]{acronym}
\modulolinenumbers[5]
\usepackage{subfloat}
\usepackage[subrefformat=parens]{subcaption}

\usepackage{booktabs}

\usepackage{amsfonts}

\journal{Astronomy and Computing}

\biboptions{authoryear}

\usepackage{xcolor}

\usepackage[labelformat=parens,labelsep=quad, skip=3pt]{caption}
\usepackage{graphicx}
\newcommand{\adjheight}{-0.2cm}

\usepackage{amsmath}

\begin{document}

\begin{frontmatter}

\title{Estimating Galaxy Redshift in Radio-Selected Datasets using Machine Learning}

\author[WSU,Data61]{Kieran J. Luken\corref{corrAuthor}\href{https://orcid.org/0000-0002-6147-693X}{\includegraphics[scale=0.5]{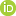}}}
\ead{k.luken@westernsydney.edu.au}
\cortext[corrAuthor]{Corresponding author}

\author[WSU,CSIRO]{Ray P. Norris\href{https://orcid.org/0000-0002-4597-1906}{\includegraphics[scale=0.5]{orcid.png}}}

\author[WSU]{Laurence A. F. Park\href{https://orcid.org/0000-0003-0201-4409}{\includegraphics[scale=0.5]{orcid.png}}}

\author[WSU,Data61]{X. Rosalind Wang\href{https://orcid.org/0000-0001-5454-6197}{\includegraphics[scale=0.5]{orcid.png}}}

\author[WSU]{Miroslav D. Filipovi\'c\href{https://orcid.org/0000-0002-4990-9288}{\includegraphics[scale=0.5]{orcid.png}}}

\address[WSU]{Western Sydney University, Locked Bag 1797, Penrith, NSW 2751, Australia}
\address[Data61]{CSIRO Data61, Epping, Australia}
\address[CSIRO]{CSIRO Astronomy and Space Sciences, Australia Telescope National Facility, PO Box 76, Epping, NSW 1710, Australia}

\begin{acronym}[ELAIS-S1]
\acro{2dfGRS}{2dF Galaxy Redshift Survey}
\acro{AGN}{Active Galactic Nucleus}
\acroplural{AGN}{Active Galactic Nucleii}
\acro{ASKAP}{Australian Square Kilometre Array Pathfinder}
\acro{ATCA}{Australia Telescope Compact Array}
\acro{ATLAS}{Australia Telescope Large Area Survey}
\acro{COSMOS}{COSMic evOlution Survey}
\acro{DES}{Dark Energy Survey}
\acro{DT}{Decision Tree}
\acro{eCDFS}{extended Chandra Deep Field South}
\acro{ELAIS-S1}{European Large Area ISO Survey--South 1}
\acro{EMU}{Evolutionary Map of the Universe}
\acro{GP}{Gaussian Process}
\acroplural{GP}{Gaussian Processes}
\acro{kNN}{$k$-Nearest Neighbours}
\acro{LMNN}{Large Margin Nearest Neighbour}
\acro{LOFAR}{LOw Frequency ARray}
\acro{LOTSS}{LOFAR Two-metre Sky Survey}
\acro{ML}{Machine Learning}
\acro{MLKR}{Metric Learning for Kernel Regression}
\acro{MOS}{Multi-Object Spectroscopy}
\acro{MSE}{Mean Square Error}
\acro{MWA}{Murchison Widefield Array}
\acro{NMAD}{Normalised Median Absolute Deviation}
\acro{NMI}{Normalised Mutual Information}
\acro{OzDES}{Australian Dark Energy Survey}
\acro{QSO}{Quasi-Stellar Object}
\acro{RF}{Random Forest}
\acro{SDSS}{Sloan Digital Sky Survey}
\acro{SED}{Spectral Energy Distribution}
\acro{SFG}{Star Forming Galaxy}
\acro{SKA}{Square Kilometre Array}
\acro{SKADS}{Square Kilometre Array Design Survey}
\acro{SST}{Spitzer Space Telescope}
\acro{SWIRE}{Spitzer Wide-Area Infrared Extragalactic Survey}
\acro{WAVES}{Wide Area Vista Extragalactic Survey}
\end{acronym}

\begin{abstract}
All-sky radio surveys are set to revolutionise the field with new discoveries. However, the vast majority of the tens of millions of radio galaxies won't have the spectroscopic redshift measurements required for a large number of science cases. Here, we evaluate techniques for estimating redshifts of galaxies from a radio-selected survey. Using a radio-selected sample with broadband photometry at infrared and optical wavelengths, we test the \ac{kNN} and \acl{RF} machine learning algorithms, testing them both in their regression and classification modes. Further, we test different distance metrics used by the \ac{kNN} algorithm, including the standard Euclidean distance, the Mahalanobis distance and a learned distance metric for both the regression mode (the \acl{MLKR} metric) and the classification mode (the \acl{LMNN} metric). We find that all regression-based modes fail on galaxies at a redshift $z > 1$. However, below this range, the \ac{kNN} algorithm using the Mahalanobis distance metric performs best, with an $\eta_{0.15}$ outlier rate of 5.85\%. In the classification mode, the \ac{kNN} algorithm using the Mahalanobis distance metric also performs best, with an $\eta_{0.15}$ outlier rate of 5.85\%, correctly placing 74\% of galaxies in the top $z > 1.02$ bin. Finally, we also tested the effect of training in one field and applying the trained algorithm to similar data from another field and found that variation across fields does not result in statistically significant differences in predicted redshifts. Importantly, we find that while we may not be able to predict a continuous value for high-redshift radio sources, we can identify the majority of them using the classification modes of existing techniques. 
\end{abstract}

\begin{keyword}
methods: analytical; techniques: photometric; galaxies: photometry; galaxies: high-redshift
\end{keyword}

\end{frontmatter}

% Uncomment the below if we want line numbers. It was a default for... Reasons?
% \linenumbers
\acresetall
\section{Introduction}

New radio telescopes are set to revolutionise the radio astronomy regime, with the \ac{EMU} project to be completed on the \acl{ASKAP} \citep[\acs{ASKAP};  \acused{ASKAP}][]{johnston_science_2007,johnston_science_2008} telescope in particular set to increase the number of known radio sources from $\sim$2.5 million \citep{norris_extragalactic_2017} to $\sim$70 million \citep{norris_emu:_2011}. 

For most aspects of science, knowledge of an astronomical source's redshift is an essential indicator of the distance and age of the source. Ideally, this redshift is measured directly using spectroscopy. However, even with modern advances --- including \ac{MOS} which can allow hundreds to thousands of redshifts to be measured at once --- deep spectroscopic surveys still fail to yield reliable redshifts from 30-60\% of measured spectra \citep{newman_spectroscopic_2015}. Currently, the \ac{SDSS} has measured $\sim$4.8 million spectroscopic redshifts --- as of the 16$^\mathrm{th}$ data release \citep{sdss_dr16}\footnote{\url{https://www.sdss.org/dr16/scope/}} --- and the \acl{2dfGRS} \citep[\acs{2dfGRS}; \acused{2dfGRS}][]{lewis_anglo-australian_2002} measured $\sim$250\,000 spectroscopic redshifts over its 5 year project\footnote{\url{http://www.2dfgrs.net/Public/Survey/statusfinal.html}}. In the future, the \ac{WAVES} survey --- expected to provide a further 2.5 million spectroscopic redshift measurements \citep{driver_wide_2016} --- will increase the number of spectroscopically measured sources. However, this will still be significantly short of the expected $\sim70$ million sources detected by the \ac{EMU} project (even if all of these newly measured galaxies were exclusively selected from the \ac{EMU} survey).

\begin{figure*}
    \centering
    \includegraphics[trim=0 0 0 0, width=\textwidth]{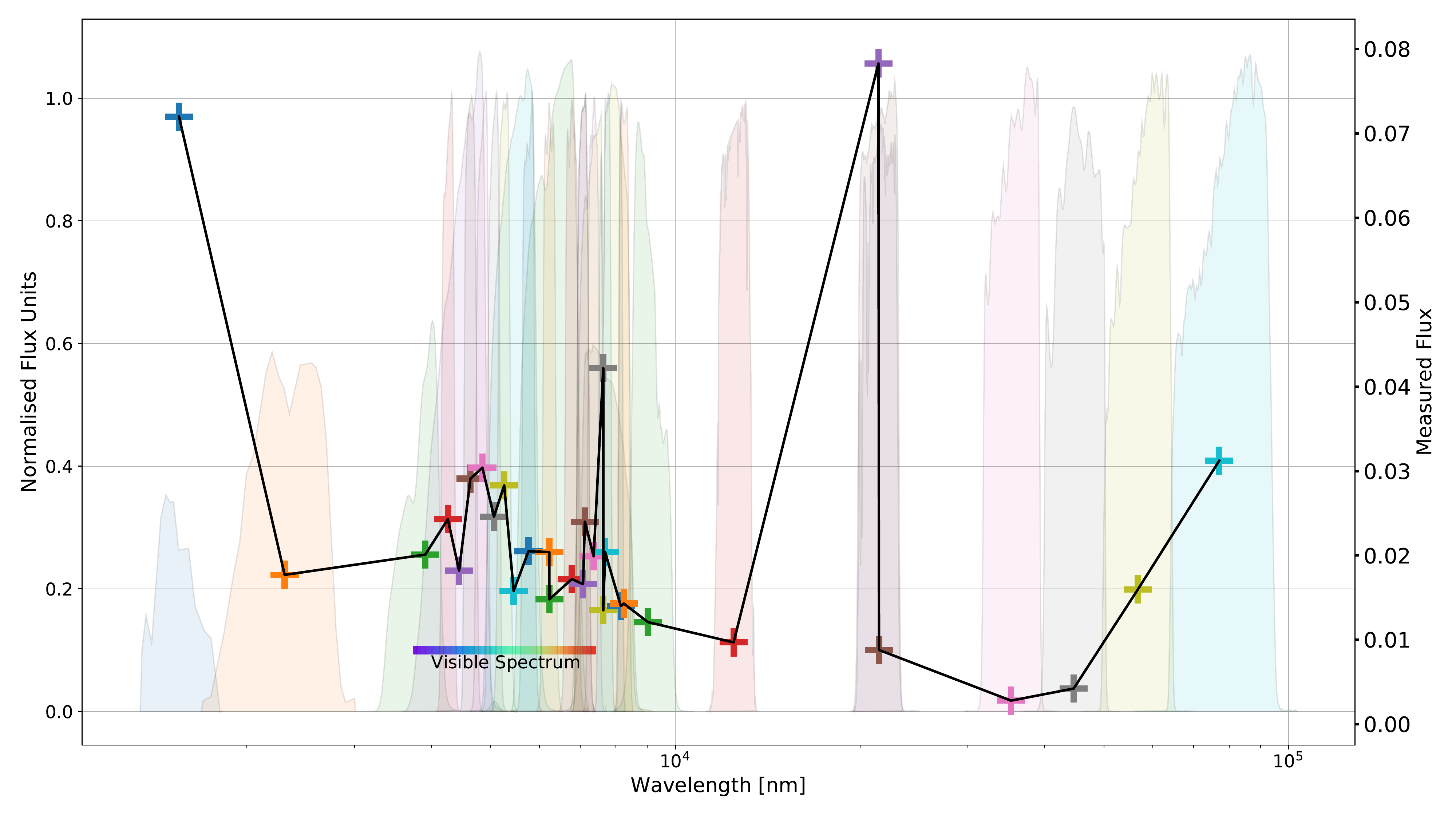}
    \caption{
        The \ac{SED} of an extragalactic source from the \ac{COSMOS} survey (Object ID: 358521). The background shows the filter coverage of the 31 filters available in the \ac{COSMOS} field (defined in \citet{ilbert_cosmos_2009}). The ``+" symbols in the foreground represent the galaxy's photometric measurements in each band, and are coloured to match the corresponding filter in the background. 
    }
    \label{fig:cosmos_photometry}
\end{figure*}

Photometric template fitting \citep{baum_photoelectric_1957} is able to estimate the redshift (hereafter $z_{photo}$) of a source, as well as ancillary data like galaxy classification. Photometric template fitting is completed by comparing the \ac{SED}, measured across as many different wavelengths as possible, to templates constructed with astrophysical knowledge, or prior examples. Figure~\ref{fig:cosmos_photometry} shows the 31 filter bands available in the \ac{COSMOS} field and an example \ac{SED} used by \citet{ilbert_cosmos_2009} to achieve redshift accuracies of $\sigma_{\Delta{z/(1+z_s)}} = 0.012$. 

\ac{ML} has also been applied to the problem, with the \ac{kNN} algorithm \citep{ball_robust_2007,ball_robust_2008,oyaizu_galaxy_2008,zhang_estimating_2013,kugler_determining_2015,cavuoti_metaphor:_2017,luken_estimating_2018,luken_neurips}, \acl{RF} \citep[\acs{RF}; \acused{RF} ][]{cavuoti_photometric_2012,cavuoti_photometric_2015,hoyle_measuring_2016,sadeh_annz2:_2016,cavuoti_metaphor:_2017,pasquet-itam_deep_2018}, and neural networks \citep{firth_estimating_2003,tagliaferri_neural_2003,collister_annz:_2004,brodwin_photometric_2006,oyaizu_galaxy_2008,hoyle_measuring_2016,sadeh_annz2:_2016,curran_1,curran_2} being among the more widely used algorithms. Some recent studies utilise \acl{GP} \citep[\acs{GP}; \acused{GP}][]{duncan_photometric_2018,duncan_photometric_2018-1,Duncan_2021}, and deep learning using the original images at different wavelengths, as opposed to the photometry extracted from the image \citep{2018A&A...609A.111D}.

\begin{figure*}
    \centering
    \includegraphics[trim=0 0 0 0, width=\textwidth]{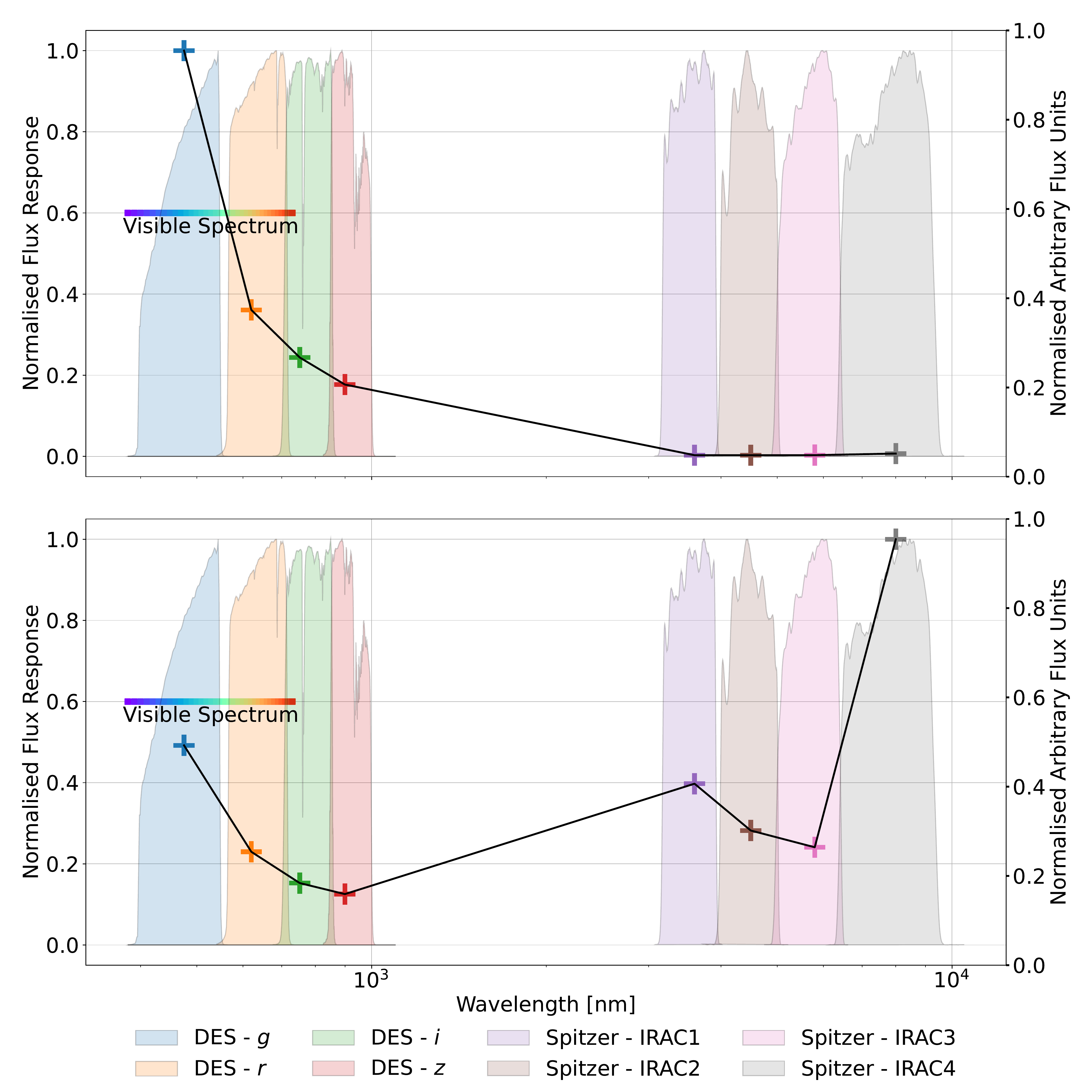}
    \caption{
    The \ac{SED} of the extragalactic source ATLAS3\_J033551.0\-283345C (top; likely \ac{AGN}) and ATLAS3\_J033519.0\-273708C (bottom; likely \ac{SFG}), in similar style to Figure~\ref{fig:cosmos_photometry}. The background shows the filter coverage used by this work, with the ``+" in the foreground representing the measured photometry at each band, taken from the data used in this study.
    }
    \label{fig:emu_photometry}
\end{figure*}

However, few of these solutions are appropriate for the large-scale radio surveys being conducted. While photometric template fitting provides a theoretically ideal solution, the quality of data required to make it highly accurate will not be available for the majority of sources in the all-sky \ac{EMU} survey, since only all-sky photometric data will be available for most sources. An example of the broadband photometry used by this work is shown in Figure~\ref{fig:emu_photometry}, which shows both a likely \acl{AGN} (\acs{AGN}; top\acused{AGN}), and \acl{SFG} (\acs{SFG}; bottom\acused{SFG}).

Additionally, \citet{norris_comparison_2018} has shown that photometric template fitting performs poorly on radio-selected data sets (which are typically dominated by \ac{AGN}), possibly because the majority of templates are unable to differentiate between emission from the \ac{AGN} and emission from the galaxy itself \citep{salvato_many_2018}.

\begin{figure}
    \centering
    \includegraphics[trim=0 0 0 0, width=0.5\textwidth]{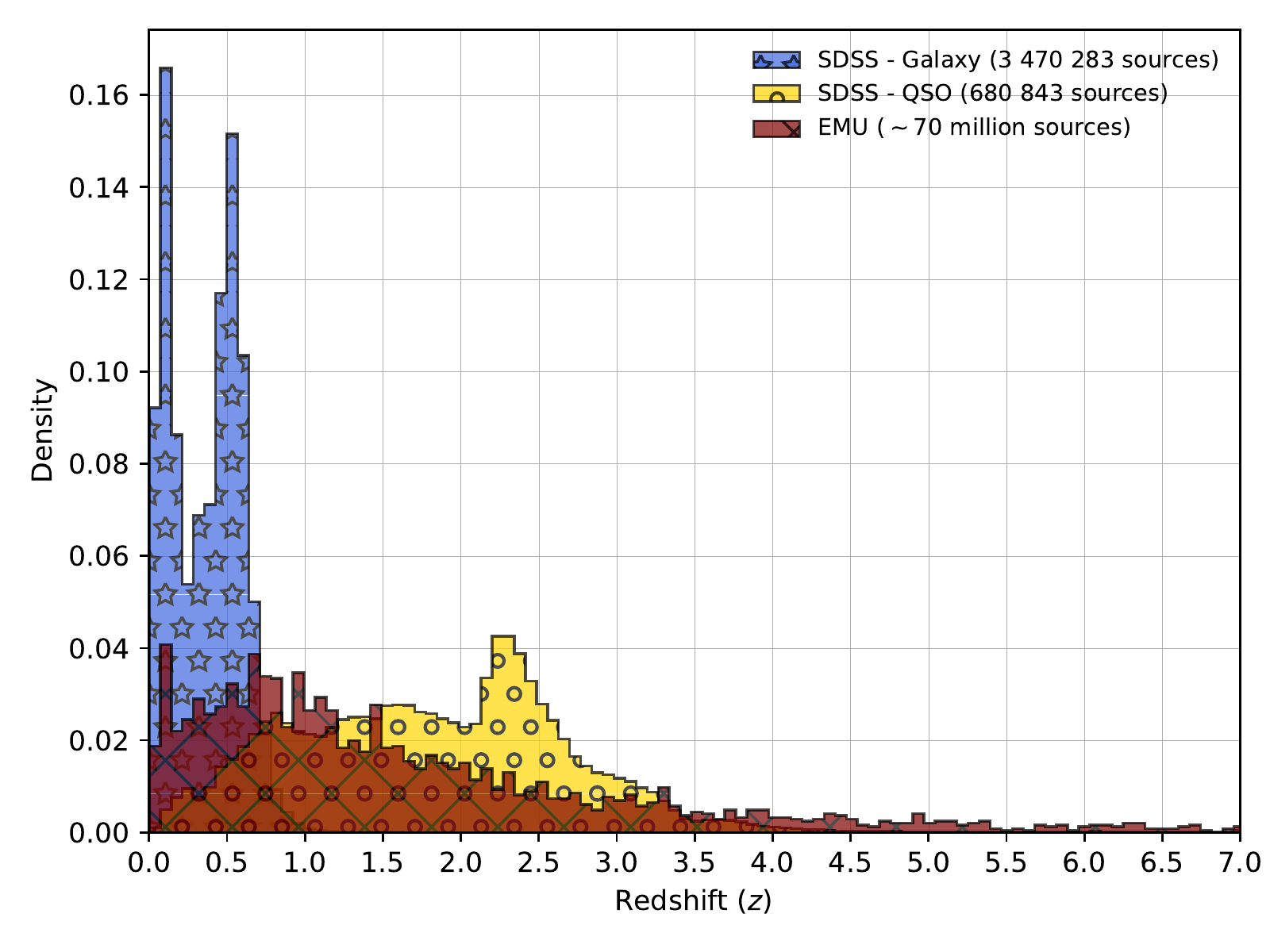}
    \caption{The normalised distributions of redshift in the \ac{SDSS} Galaxy and \ac{QSO} surveys, compared with the expected redshift distribution of the \ac{EMU} survey, modelled by the \acl{SKADS} \citep[\acused{SKADS}\acs{SKADS};][]{SKADS}.}
    \label{fig:emu_histogram}
\end{figure}

The \ac{ML} based methods have mainly focused on optically-selected datasets, with most work drawing on the \ac{SDSS} photometric and spectroscopic samples, often restricting the redshift range to $z < 1$, or using datasets containing only one type of object --- e.g. the \ac{SDSS} Galaxy or \ac{QSO} catalogs. While this kind of testing is entirely appropriate for non-radio selected samples, \citet{norris_comparison_2018} have shown that assumptions in the optical regime may not be valid in the radio regime, for two reasons. First, the redshift distribution of radio-selected sources is quite different from that of  optically-selected sources, as shown in  Figure~\ref{fig:emu_histogram}. Second, radio-selected sources are often dominated by a radio-loud AGN which is poorly represented in optically-selected templates and training sets. 

Finally, we should note that in support of the science goals of the \ac{EMU} Project, we aim to minimise the number of estimations that catastrophically fail, rather than minimise for accuracy as most other works do. 

The overall contributions of this study include:
\begin{itemize}
    \item An in-depth study of the simple \acl{kNN} algorithm, including the investigation of typically ignored distance metrics for the estimation of radio galaxy redshift;
    \item A comparison with the widely used \acl{RF} algorithm;
    \item An analysis of whether similarly observed fields can be used as training samples for alternate fields; 
    \item An exploration into the effectiveness of regression vs classification modes in estimating high redshift galaxies, given the highly unbalanced nature of the training sample (noting that while classification modes might be able to identify high-redshift galaxies, it will not be able to estimate its actual redshift). 
\end{itemize}

\subsection{Formal Problem Statement}

In this article, we investigate the problem of modelling the redshift of a source with respect to known measurements of the source. This can be a regression problem, where we attempt to model the function $r_i = f(\vec{x}_i)$, such that the redshift of source $i$, $r_i \in \mathbb{R}^+$ and $\vec{x}_i$ the known measurements of source $i$ are obtained from a catalogue with various domains. We can also model the redshift of a source as $c_i = g(\vec{x}_i)$, where $c_i$ is a class representing a specific domain of redshift.
In the following section we describe the catalogues containing the known source information $\vec{x}_i$ and the variables that are used for this representation.

\section{Data}
\label{sec:data}

This work uses data from the \acl{ATLAS} \citep[\acused{ATLAS}\acs{ATLAS};][]{norris_deep_2006,franzen_atlas_2015} radio continuum catalogue, providing a 1.4~GHz flux density measurement of 4780 unique sources. The \ac{ATLAS} survey covers two regions of the sky -- the \ac{eCDFS} and the \ac{ELAIS-S1}, both to a depth of $\sim 15 \mu\textrm{Jy}$, and was completed as a first-look at what the \ac{EMU} survey may provide. This survey was cross-matched by \citet{ATLAS_Swan} with the \acl{OzDES} \citep[\acused{OzDES}\acs{OzDES};][]{OZDES_1,OZDES_2,OZDES_3}, \acl{DES} \citep[\acused{DES}\acs{DES};][]{dark_energy_survey_collaboration_dark_2016} and \acl{SWIRE} \citep[\acused{SWIRE}\acs{SWIRE};][]{lonsdale_swire:_2003} surveys, providing spectroscopic redshift measurements, \textit{g}, \textit{r}, \textit{i} and \textit{z} optical magnitudes, and 3.6, 4.5, 5.4 and 8.0 $\mu$m infrared flux measurements respectively.

\begin{figure}
    \centering
    \includegraphics[trim=0 0 0 0, width=0.5\textwidth]{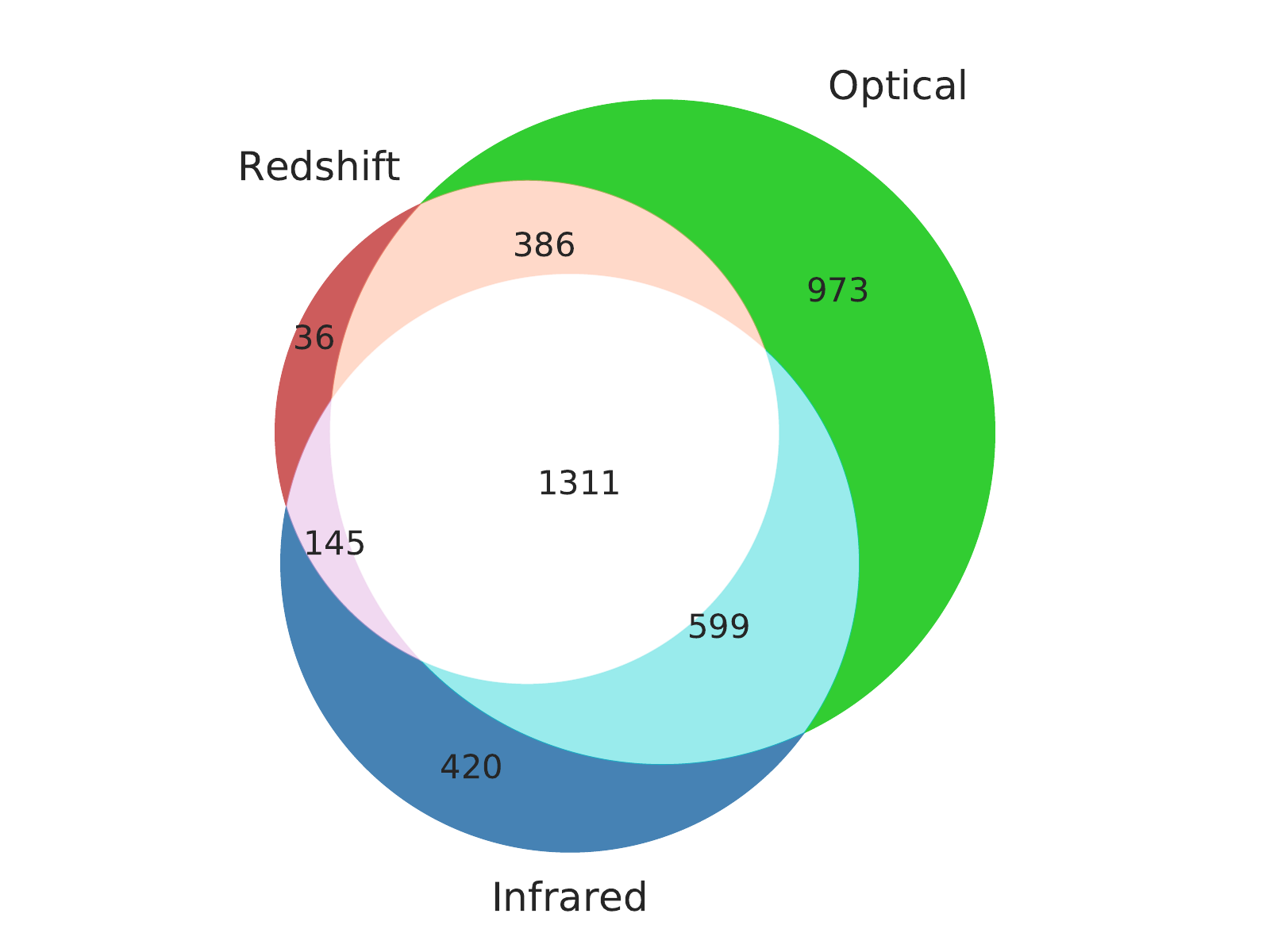}
    \caption{A Venn diagram of the number of radio sources in the \ac{ATLAS} dataset, with a spectroscopic redshift (provided by the \ac{OzDES}), optical magnitudes at $g$, $r$, $i$, and $z$ bands (provided by the \ac{DES}) and  infrared fluxes at 3.6, 4.5, 5.8 and 8.0 $\mu$m (provided by the \ac{SWIRE} survey)
    }
    \label{fig:dataset_venn}
\end{figure}

\begin{figure}
    \centering
    \includegraphics[trim=0 0 0 0, width=0.5\textwidth]{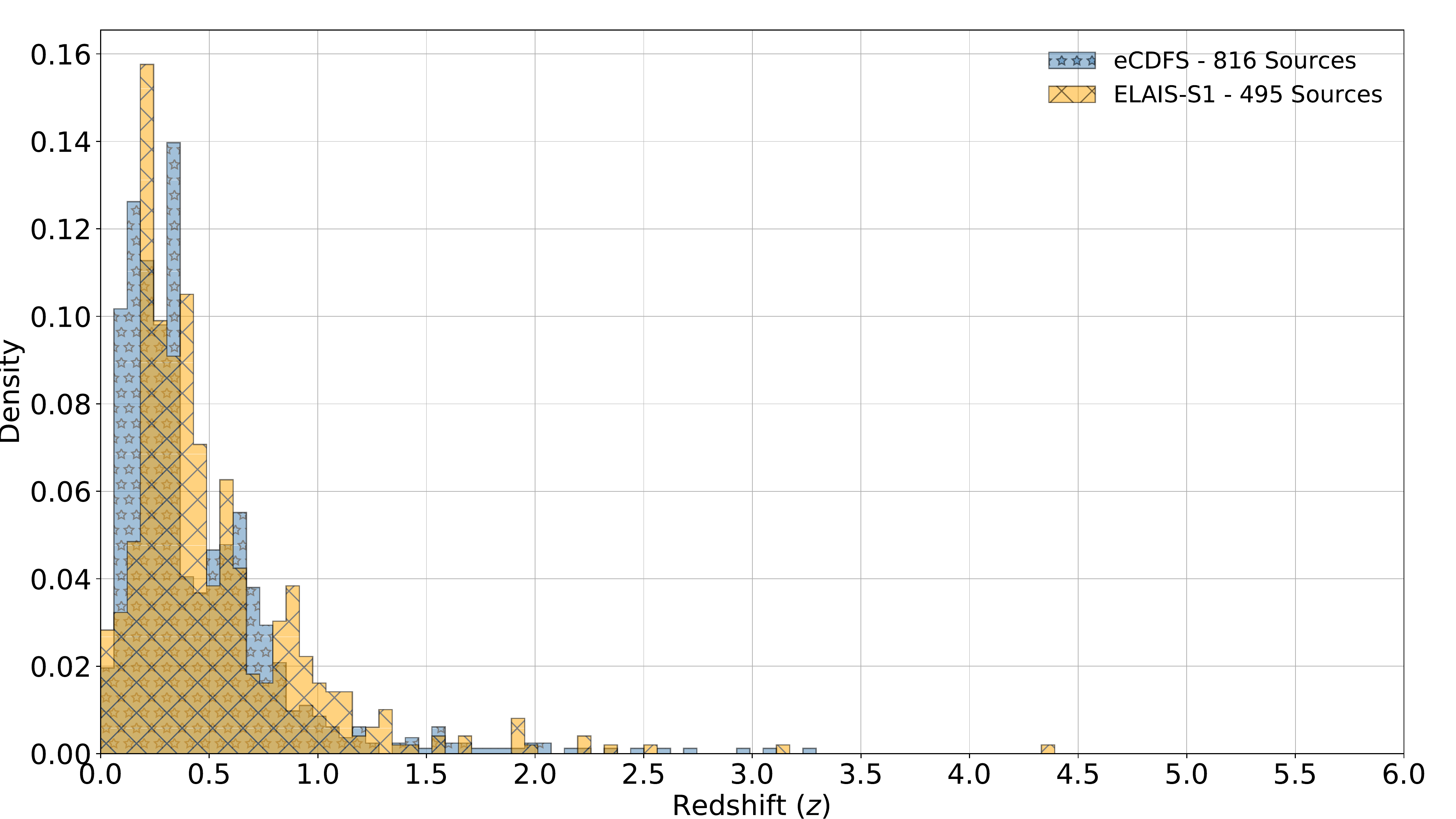}
    \caption{The redshift distribution of sources in the \ac{ATLAS} dataset, broken down by field (\ac{ELAIS-S1} and \ac{eCDFS}). The overall mean redshift is $z = 0.47$, \ac{ELAIS-S1} mean is $z = 0.51$ and \ac{eCDFS} mean is $z = 0.44$}
    \label{fig:atlas_redshift}
\end{figure}

This work used only those \ac{ATLAS} sources with measured photometry in all provided optical and infrared bands, creating a final dataset containing 1311 sources with complete photometric coverage. Specifically, in both fields used in this work, we used the \textit{g}, \textit{r}, \textit{i} and \textit{z} optical magnitudes from \ac{DES}, and 3.6, 4.5, 5.4 and 8.0 $\mu$m infrared flux measurements from the \ac{SWIRE} survey, as shown in Figure~\ref{fig:dataset_venn}. The redshift distribution of the final collated dataset is presented in Figure~\ref{fig:atlas_redshift}.

To prevent the different methods tested from being dominated by single features with wide variation, all features were standardised using Equation~\ref{eqn:standardisation}, setting the feature mean to 0, and the feature variance to unit variance. 

\begin{equation}
    z_i = \frac{x_i - \bar{x}}{s_x},
    \label{eqn:standardisation}
\end{equation}
where $\bar{x}$ is the mean of the sample of variable $x$ and $s_x$ is its standard deviation.

During testing, we also examined the use of ``colours'' rather than optical magnitudes, and taking the log of the radio and infrared fluxes to better distribute the data. Colours are typically used instead of magnitudes to remove the brightness- and redshift-dependent nature of magnitudes, replacing them by the difference between the magnitudes that is dependent only on the \ac{SED}.

\subsection{Data Partitioning}

All machine learning methods require the main dataset to be partitioned into multiple subsets, with the subsets set aside for training the model, validating hyperparameters, and testing the model. In this work, we split our data into two datasets -- training and testing, with the hyper-parameters being validated using $k$-fold cross validation on the training set. 
We partition our data differently for the three tests we complete:

\begin{itemize}

\item Typical random split, with 70\% of the data split off as the training set, leaving the remaining 30\% as the test set. 

\item The entire \ac{ELAIS-S1} field as the training set, leaving the \ac{eCDFS} field as the test set. 

\item The entire \ac{eCDFS} field as the training set, leaving the \ac{ELAIS-S1} field as the test set. 

\end{itemize}
In addition to the above partitioning, in order to facilitate the use of the classification modes of our algorithms tested, we quantise the redshift values into 15 redshift bins (defined in Table~\ref{table:classification_bounds}), with equal numbers of sources in each in order to be able to predict a uniform distribution with using a matching distribution. The median of the spectroscopic redshifts within the bin is chosen as the redshift to predict.

\begin{table}
\centering
\caption{Details of the bins used for the classifications tests, outlining the lower edge, upper edge and predicted value of each bin (where the predicted value is the median of the spectroscopic redshifts from within the bin).}
\label{table:classification_bounds}
\begin{tabular}{cccc}
\hline 
   Bin & Lower & Upper & Median \\
   Number & Bound & Bound & Value \\
\hline 
    1 & 0 & 0.10 & $< 0.1$ \\
    2 & 0.10 & 0.15 & 0.12 \\
    3 & 0.15 & 0.19 & 0.17 \\
    4 & 0.19 & 0.22 & 0.21 \\
    5 & 0.22 & 0.26 & 0.24 \\
    6 & 0.26 & 0.29 & 0.27 \\
    7 & 0.29 & 0.32 & 0.31 \\
    8 & 0.32 & 0.35 & 0.34 \\
    9 & 0.35 & 0.41 & 0.38 \\
    10 & 0.41 & 0.50 & 0.46 \\
    11 & 0.50 & 0.58 & 0.54 \\
    12 & 0.58 & 0.66 & 0.62 \\
    13 & 0.66 & 0.80 & 0.73 \\
    14 & 0.80 & 1.02 & 0.91 \\
    15 & 1.02 & 4.33 & $> 1.02$ \\
\end{tabular}
\end{table}

\section{Methods}

\subsection{\ac{kNN}}

\begin{figure}
    \centering
    \includegraphics[trim=0 0 0 0, width=0.5\textwidth]{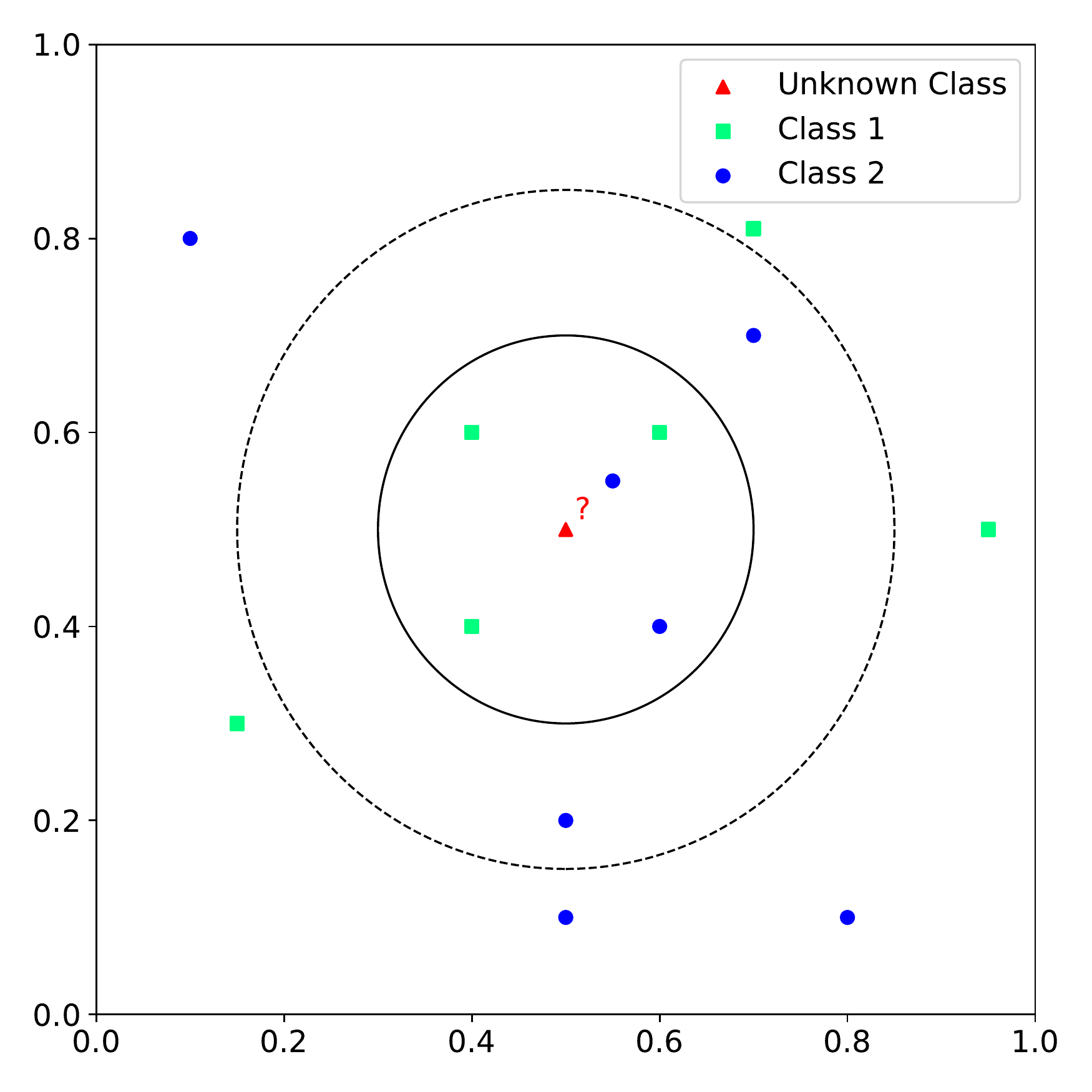}
    \caption{A pictorial representation of the \acl{kNN} algorithm. If $k_N$ is chosen to be 5 (denoted by the solid circle), by taking the mode of the $k_N$ nearest neighbours, the red triangle would be assigned a class of Green Square. If the value of $k_N$ was changed to 7 (denoted by the dashed circle), the mode of the $k_N$ nearest neighbours would be the Blue Circle. Modified from: \citet{ajanki_antti_example_2007}}
    \label{fig:knn}
\end{figure}

The \ac{kNN} algorithm \citep{cover_nearest_1967} computes a similarity matrix between all sources based on the catalogue's photometric measurements, and a given distance metric (the set of distance metrics tested in this work are explained in Subsections~\ref{subsubsec:euclidean}, \ref{subsubsec:mahal}, and \ref{subsubsec:learned_metric}). Once the similarity matrix is constructed, the \ac{kNN} algorithm finds the $k$ (Hereafter $k_N$) most similar sources with measured redshifts (where $k_N$ is optimised using cross validation), and takes either the mean value (for regression) or the mode class (for classification) of the sources as the estimated redshift for each source. A simple illustration of the \ac{kNN} algorithm is shown in Figure~\ref{fig:knn}

The value of $k_N$ used in the \ac{kNN} in this work is optimised using $k$-fold (where $k$ is hereafter $k_f$ and is set to 10 for this work) cross-validation. $k_f$-fold cross-validation randomly splits the data into $k_f$ roughly even groups, iterating through using $k_f-1$ groups as the training set and validating on the remaining group, until every group has been used in testing, with the average error used as the error for that value of $k_f$. This work tested all integer values of $k_N$ for the \ac{kNN} algorithm between 2 and 20 ($k_N \in \{2,...,20\}$).

The \ac{kNN} algorithm requires a metric to determine which of the observations are neighbours to a given observation. In the next sections we present the set of metrics investigated in the experiment.

\subsubsection{Euclidean Distance}
\label{subsubsec:euclidean}

Euclidean Distance is the simplest, and most widely used distance metric in literature, defined in Equation~\ref{eqn:euclidean}:

\begin{equation} 
    \label{eqn:euclidean}
    d(\vec{p},\vec{q}) = \sqrt{(\vec{p} - \vec{q})^\mathrm{T}(\vec{p} - \vec{q})}, 
\end{equation} 
where $d(\vec{p},\vec{q})$ is the Euclidean distance between two feature vectors $\vec{p}$ and $\vec{q}$. In this case the vectors contain the measured photometry of two galaxies.

\subsubsection{Mahalanobis Distance}
\label{subsubsec:mahal}

The Mahalanobis distance metric \citep{mahalanobis1936generalized} normalises the variance and covariance of the input features by transforming the features using the inverse of the covariance matrix. For uncorrelated input features, the Mahalanobis distance is equal to the scaled Euclidean distance, but for correlated features, it generalises the idea of Euclidean distance to take account of the covariance.  The Mahalanobis distance is defined in Equation~\ref{eqn:mahalanobis}:

\begin{equation} 
	\label{eqn:mahalanobis} 
    d(\vec{p},\vec{q}) = \sqrt{(\vec{p} - \vec{q})^\mathrm{T}S^{-1}(\vec{p} - \vec{q})}, 
\end{equation}
where $d(\vec{p},\vec{q})$ is the Mahalanobis distance between two feature vectors $\vec{p}$ and $\vec{q}$, and $S$ is the covariance matrix.  

\subsubsection{Learned Distance Metrics}
\label{subsubsec:learned_metric}

Conceptually, for best results the distance metric used should take into consideration the shape and structure of the data. Towards this end, we can generalise Equations~\ref{eqn:euclidean}, and \ref{eqn:mahalanobis} to Equation~\ref{eqn:metricLearn}, noting that the $\mathbf{M}$ matrix can be any positive semi-definite matrix. The Identity matrix is used for Euclidean Distance, and the $S^{-1}$ matrix is used in the Mahalanobis. We can, however, go one step further, and attempt to learn an $\mathbf{M}$ matrix that can better warp the feature space so that observations with similar redshift are measured as close, while observations with different redshift are measured as distant.

\begin{equation}
    \label{eqn:metricLearn}
    d(\vec{p},\vec{q})=\sqrt{(\vec{p}-\vec{q})^\mathrm{T}\mathbf{M}(\vec{p}-\vec{q})},
\end{equation}
where $d(\vec{p},\vec{q})$ is the distance between two feature vectors  $\vec{p}$ and $\vec{q}$.

For our regression tests, we used the \ac{MLKR} distance metric, which performs a supervised Principle Component Analysis \citep{weinberger_metric_2007}. The \ac{MLKR} distance metric begins by decomposing the $\mathbf{M}$ matrix from Equation~\ref{eqn:metricLearn} using Equation~\ref{eqn:mlkr_decomposed}:

\begin{equation}
    \label{eqn:mlkr_decomposed}
    \mathbf{M} = \mathbf{A}^\mathrm{T} \mathbf{A} 
\end{equation}

Using Equation~\ref{eqn:mlkr_decomposed}, Equation~\ref{eqn:metricLearn} can be expressed as the modified Euclidean Distance metric in Equation~\ref{eqn:mlkr_euclidean}:

\begin{equation}
    \label{eqn:mlkr_euclidean}
    d(\vec{x}_i,\vec{x}_j)=||\textbf{A}(\vec{x}_i-\vec{x}_j)||^2
\end{equation}

Matrix $\mathbf{A}$ is optimised using Gradient Descent, using Equation~\ref{eqn:mlkr}:

\begin{equation}
    \label{eqn:mlkr}
    \frac{\partial \mathcal{L}}{\partial \mathbf{A}} = 4\mathbf{A}\sum_i (\hat{y}_i - y_i) \sum_j(\hat{y}_j - y_j) k_{ij}\vec{x}_{ij} \vec{x}_{ij}^\mathrm{T},
\end{equation}
where $\hat{y}_i$ is defined in Equation~\ref{eqn:mlkr_yhat}, $k_{ij}$ is a Gaussian kernel defined in Equation~\ref{eqn:mlkr_gaussian}, and $\vec{x}_{ij} = (y_i - \hat{y}_i)^2$.

\begin{equation}
    \label{eqn:mlkr_yhat}
    \hat{y}_i = \frac{\sum_{y \neq i} y_jk_{ij}}{\sum_{j \neq k} k_{ij}}
\end{equation}

\begin{equation}
    \label{eqn:mlkr_gaussian}
    k_{ij} = \frac{1}{\sigma \sqrt{2\pi}}e^{-\frac{d(\vec{x}_i,\vec{x}_j)}{\sigma^2}}
\end{equation}

The loss function being minimised is a simple squared difference, defined in Equation~\ref{eqn:mlkr_loss}:

\begin{equation}
    \label{eqn:mlkr_loss}
    \mathcal{L} = \sum_i (y_i - \hat{y}_i)^2
\end{equation}

For our Classification tests, the \acl{LMNN} \citep[\acused{LMNN}\acs{LMNN}; ][]{weinberger_distance_2006} learned distance metric was used. The \ac{LMNN} distance metric finds a transformation for the data that maximises the distance between different classes, and minimises the distance between similar classes. The loss function the \ac{LMNN} algorithm optimises is defined in Equation~\ref{eqn:lmnn}:

\begin{equation}
    \label{eqn:lmnn}
    \begin{split}
        \epsilon (\mathbf{L}) = \sum_{ij} \eta_{ij} || \mathbf{L}(\vec{x}_i - \vec{x}_j) ||^2 
        \\ + c \sum_{ijl} \eta_{ij}(1-y_{il}) [ 1 + 
        \\ || \mathbf{L}(\vec{x}_i - \vec{x}_j) ||^2 - 
        \\ ||\mathbf{L}(\vec{x}_i - \vec{x}_l) ||^2]_+ ,
    \end{split}
\end{equation}
where $i$, $j$, and $l$ are individual galaxy feature vectors, $\eta_{ij} \in \{0,1\}$ describes whether $\vec{x}_i$ is a target of $\vec{x}_j$, $c$ is a positive constant typically chosen through cross-validation and $[z]_+ = \mathrm{max}(z,0)$ -- a hinge function. 

\begin{figure}
    \centering
    \includegraphics[trim=0 0 0 0, width=0.45\textwidth]{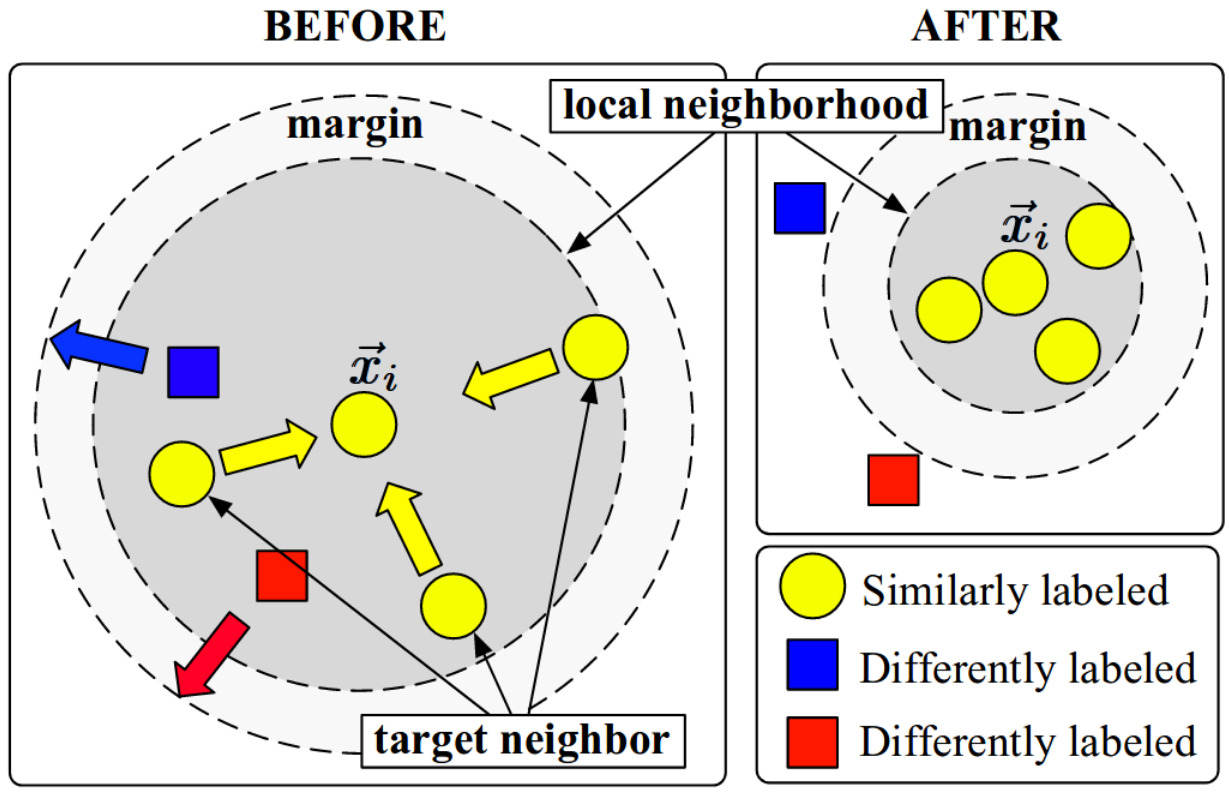}
    \caption{An example of the transformation found by the \acl{LMNN} algorithm. The distance between similar classes is minimised, and the distance between different classes is maximised. }
    \label{fig:lmnn}
\end{figure}

An example of the transformation the \ac{LMNN} algorithm attempts is in Figure~\ref{fig:lmnn}, using three neighbours.

\subsection{\acl{RF}}

For comparison, we have contrasted our results with the popular and well-used \ac{RF} algorithm. \acp{RF} are constructed from bootstrapped \acp{DT} -- an algorithm that partitions the data space so that it can be explored using a tree, shown in Figure~\ref{fig:dt} \citep{morgan_problems_1963,quinlan_simplifying_1987}. Each \ac{DT} finds the impurity at each node (defined in Equation~\ref{eqn:rf_impurity}) to determine the best features and values to split the data on:

\begin{figure}
    \centering
    \includegraphics[trim=0 0 0 0, width=0.45\textwidth]{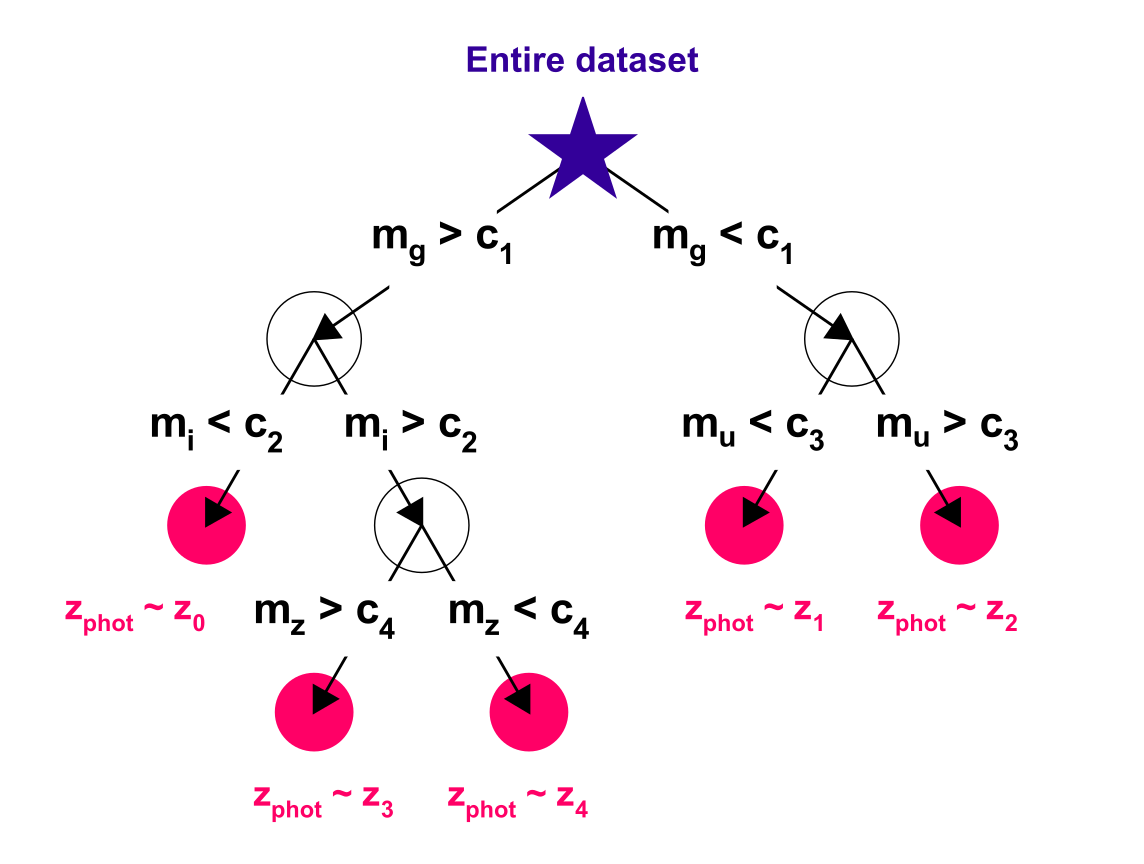}
    \caption{An example of a \acl{DT}. Starting at the top with the entire dataset, questions are  asked of the data until an outcome is clearly identified.}
    \label{fig:dt}
\end{figure}

\begin{equation}
    \label{eqn:rf_impurity}
    G(Q, \theta) = \frac{n_{left}}{N_m} H(Q_{left}(\theta)) + \frac{n_{right}}{N_m} H(Q_{right}(\theta)) ,
\end{equation}
where $Q$ is the data at node $m$, $\theta$ is a subset of data, $N_m$ is the number of objects at node $m$, $n_{left}$ and $n_{right}$ are the numbers of objects on the left and right sides of the split, $Q_{left}$ and $Q_{right}$ are the objects on the left and right sides of the split, and the $H$ function is a impurity function that differs between classification and regression. For Regression, the Mean Square Error is used (defined in Equation~\ref{eqn:rf_mse}), whereas Classification often uses the Gini Impurity (defined in Equation~\ref{eqn:rf_gini}).

\begin{equation}
    \label{eqn:rf_mse}
    H(X_m) = \frac{1}{N_m} \sum_{i \in N_m} (y_i - \left[\frac{1}{N_m} \sum_{j \in N_m} y_j\right] )^2,
\end{equation}
where $N_m$ is the number of objects at node $m$ and $y_i$ and $y_j$ are the response variables.

\begin{equation}
    \label{eqn:rf_gini}
    H(X_m) = \sum_k p_{mk} (1 - p_{mk}),
\end{equation}
where $p_{mk}$ is the proportion of split $m$ that are class $k$, defined formally in Equation~\ref{eqn:rf_gini_pmk}:

\begin{equation}
    \label{eqn:rf_gini_pmk}
    p_{mk} = \frac{1}{N_m} \sum_{x_i \in R_m} I(y_i = k),
\end{equation}
where $I$ is the indicator function, identifying the correct classifications.

\subsection{Error Metrics}
\label{subsec:Error}

In order to evaluate each of the machine learning models, the prediction error of each model will be assessed. In this section, the set of statistics that measure the error of the models are presented. The primary error metric compared in this work is the $\eta_{0.15}$ outlier rate, defined in Equation~\ref{eqn:outlier}:

\begin{equation}
    \label{eqn:outlier}
    \eta_{0.15} = \frac{\text{count}(|\Delta z| > 0.15 \times (1 + z_{spec}) )}{\text{Number Of Sources}},
\end{equation} 
where $\Delta z = z_{spec} - z_{photo}$. The $\eta_{0.15}$ outlier rate is a percentage representing the number of `catastrophic failures' (the percentage of galaxies that have a residual greater than 0.15, scaled with redshift), and is commonly found in literature \citep{ilbert_cosmos_2009,salvato_photometric_2009,salvato_dissecting_2011,cavuoti_photometric_2012,zitlau_stacking_2016,cavuoti_metaphor:_2017,jones_analysis_2017,mountrichas_estimating_2017,luken_estimating_2018,norris_comparison_2018}.

We also provide a secondary $\eta_{2\sigma}$ outlier rate in order to provide a statistically sound comparison, defined in Equation~\ref{eqn:outlier_2sigma}:

\begin{equation}
    \label{eqn:outlier_2sigma}
    \eta_{2\sigma} = \frac{\text{count}(|\Delta z| > 2 \sigma )}{\text{Number Of Sources}} \times 100 ,
\end{equation} 
where $\Delta z = z_{spec} - z_{photo}$, and $\sigma$ is the standard deviation of the estimated response, defined in Equation~\ref{eqn:outlier_sigma}:

\begin{equation}
    \label{eqn:outlier_sigma}
    \sigma = \sqrt{\frac{1}{N} \sum_{i=1}(y_i - \hat{y})^2},
\end{equation}
where $N$ is the number of observations, $y_i$ is an observation, and $\hat{y}_i$ is the estimated observation.

As this work is presenting both regression and classification modes of the given algorithms, there are some error metrics that suit regression, and some that suit classification. These metrics are defined in Subsection~\ref{subsec:Error_regress} and \ref{subsec:Error_class} respectively. 

\subsubsection{Regression Error Metrics}
\label{subsec:Error_regress}

For the regression tests, three additional error metrics were compared. The first is the \ac{NMAD}. The \ac{NMAD} is a similar measure to the standard deviation, and is generally used for non-Gaussian distributions. It is more robust than the standard deviation, as it takes the median of the residuals, improving the resilience to outliers -- an issue that can be prominent in redshift estimation due to the highly unbalanced data sets used, and is defined in Equation~\ref{eqn:nmad}:
\begin{equation}
    \label{eqn:nmad}
    \sigma_{\mathrm{NMAD}} = 1.4826 \times (\mathrm{median}(|X_i - \mathrm{median}(X)|),
\end{equation}
where $\mathrm{NMAD}$ is the Normalised Median Absolute Deviation, $X$ is a vector of residuals from which $X_i$ is taken. 

The $R^2$ Coefficient of Determination is the second regression-based error metric, and is the proportion of variance explained by the model, The $R^2$ is defined in Equation~\ref{eqn:r2}:

\begin{equation}
    \label{eqn:r2}
    R^2 = 1 - \frac{\sum_{i} (y_j - \hat{y}_i)^2}{\sum_{i} (y_i - \bar{y})^2},
\end{equation}
where $y_i$ is a response variable, $\hat{y}$ is the corresponding estimated response, and $\bar{y}$ is the mean of the response variables.

Finally, the \ac{MSE} is a direct measure of the error produced by the model, with the lower the value the better, and is defined in Equation~\ref{eqn:mse}:

\begin{equation}
    \label{eqn:mse}
    \operatorname{MSE} =\frac {1}{N} \sum _{i=1}^{N}(y_{i}-\hat {y_{i}})^{2},
\end{equation}
where $N$ is the number of observations, $y_i$ is the measured response, and $\hat{y}_i$ is the estimated response.

\subsubsection{Classification Error Metrics}
\label{subsec:Error_class}

Traditionally in \ac{ML} classification settings, the Accuracy (defined in Equation~\ref{eqn:accuracy}), Precision, Recall and F1 score are reported. We also investigated the \acl{NMI}, which suggests how dependent one set of data is upon another. We do not report the Precision, Recall, F1 Score and \acl{NMI}, however, as we found they provided no additional information.

It is important to note that the main error metric being compared and minimised in both the regression and classification tests is the $\eta_{0.15}$ outlier rate. As this outlier rate is explicitly accepting of a level of inaccuracy that scales with redshift, there is an inherent acceptance of some level of leakage between neighbouring classes during classification tests.

This acceptance of leakage means that models may present with high error rates due to rigid correct/incorrect classifications, and yet may still be acceptable models for this process. 

\begin{equation}
    \label{eqn:accuracy}
    \mathrm{Accuracy}(y, \hat{y}) = \frac{1}{n} \sum_{i=0}^{n-1} I(\hat{y}_i = y_i),
\end{equation}
where $y$ is the measured response, $\hat{y}$ is the predicted response, $n$ is the number of samples, and $I$ is an indicator function, indicating the cases where the predicted response matched the measured response.

\subsection{Statistical Significance}
\label{subsec:stats}
Analysis of Variance (ANOVA) tests were used in this study to test for statistical significance. This allows us to test whether the changes in model correlate to a statistically significant change in estimated redshift, by testing to see if the means of two or more populations (in this case, experiments with different models) differ. 

In all cases, the tests were run in a one-vs-many scenario, with the one being tested as our best performing metric in order to determine whether our best performing result is statistically significant.  

\subsection{Software}

This work makes use of the Scikit-learn\footnote{\url{https://scikit-learn.org/}} Python package \citep{pedregosa_scikit-learn:_2011} for the implementation of the \ac{RF} and \ac{kNN} algorithms, as well as the Euclidean and Mahalanobis distance metrics. We made use of the PyLMNN\footnote{\url{https://pypi.org/project/PyLMNN/}} package for the \ac{LMNN} distance metric, and the metric-learn\footnote{\url{http://contrib.scikit-learn.org/metric-learn/generated/metric_learn.MLKR.html}} package for the \ac{MLKR} distance metric. The code and data used in this work is available on Github\footnote{\url{https://github.com/kluken/Redshift-kNN-2021}}. 

\section{Results}
\label{sec:results}

Given the number of different algorithms, distance metrics and datasets tested in this work, we have assigned each combination a Test ID, defined in Table~\ref{table:RegressionTestID} for Regression-based tests, and Table~\ref{table:ClassificationTestID} for Classification-based tests. Each test ID is made up of 4 characters, with the first character (R/C) representing whether the test is a regression- or classification-based test, the 2nd and 3rd characters (Eu/Ma/ML/Rf) representing the method used for estimation (Euclidean distance, Malanobis distance, a Learned distance metric, or Random Forest), and the final character (1/2/3) representing the training set used (1 = Random, 2 = \ac{ELAIS-S1}, 3 = \ac{eCDFS}).

As discussed in Section~\ref{sec:data}, we tested both statistical standardisation of input photometry, compared with the use of astrophysically derived ``colours'', and taking the log of radio and infrared data to better distribute the data. In both cases, however, the result was very similar to the standardised dataset, and hence is the only value quoted.

\begin{table}
	\centering
    \caption{Description of the regression based tests conducted. The Test ID is used to reference the results in Table~\ref{table:regressionResults}. The Method is the \ac{ML} algorithm used in the test. The Distance Metric is the distance metric used in the \ac{kNN} algorithm (and hence is empty for the \ac{RF} algorithm). The Training Set is the method of choosing the training set for the \ac{ML} algorithm. }
    \label{table:RegressionTestID}
    \begin{tabular}{cccc}
    	\toprule
    Test ID & Method & Distance Metric & Training Set \\    
    \midrule
    REu1 & \ac{kNN} & Euclidean & Random \\
    REu2 & \ac{kNN} & Euclidean & \ac{ELAIS-S1} \\
    REu3 & \ac{kNN} & Euclidean & \ac{eCDFS} \\
    RMa1 & \ac{kNN} & Mahalanobis & Random \\
    RMa2 & \ac{kNN} & Mahalanobis  & \ac{ELAIS-S1} \\
    RMa3 & \ac{kNN} & Mahalanobis  & \ac{eCDFS} \\
    RML1 & \ac{kNN} & \ac{MLKR}  & Random \\
    RML2 & \ac{kNN} & \ac{MLKR}  & \ac{ELAIS-S1} \\
    RML3 & \ac{kNN} & \ac{MLKR}  & \ac{eCDFS} \\
    RRf1 & \ac{RF} & -- & Random \\
    RRf2 & \ac{RF} & -- & \ac{ELAIS-S1} \\
    RRf3 & \ac{RF} & -- & \ac{eCDFS} \\
	\bottomrule
	\end{tabular}
\end{table}

\begin{table}
	\centering
    \caption{Description of the classification based tests conducted. The Test ID is used to reference the results in Table~\ref{table:classification_results}. The Method is the \ac{ML} algorithm used in the test. The Distance Metric is the distance metric used in the \ac{kNN} algorithm (and hence is empty for the \ac{RF} algorithm). The Training Set is the method of choosing the training set for the \ac{ML} algorithm. }
    \label{table:ClassificationTestID}
    \begin{tabular}{cccc}
    	\toprule
    Test ID & Method & Distance Metric & Training Set  \\
    \midrule
    CEu1 & \ac{kNN} & Euclidean & Random \\
    CEu2 & \ac{kNN} & Euclidean & \ac{ELAIS-S1} \\
    CEu3 & \ac{kNN} & Euclidean & \ac{eCDFS} \\
    CMa1 & \ac{kNN} & Mahalanobis & Random \\
    CMa2 & \ac{kNN} & Mahalanobis & \ac{ELAIS-S1} \\
    CMa3 & \ac{kNN} & Mahalanobis & \ac{eCDFS} \\
    CML1 & \ac{kNN} & \ac{LMNN} & Random \\
    CML2 & \ac{kNN} & \ac{LMNN} & \ac{ELAIS-S1} \\
    CML3 & \ac{kNN} & \ac{LMNN} & \ac{eCDFS} \\
    CRf1 & \ac{RF} & -- & Random \\
    CRf2 & \ac{RF} & -- & \ac{ELAIS-S1} \\
    CRf3 & \ac{RF} & -- & \ac{eCDFS} \\
	\bottomrule
	\end{tabular}

\end{table}

\subsection{Regression}

\begin{figure*}[ht!]
    \centering
	\begin{subfigure}[b]{0.32\textwidth}
		\caption{} \vspace{\adjheight}
		\includegraphics[width = \textwidth]{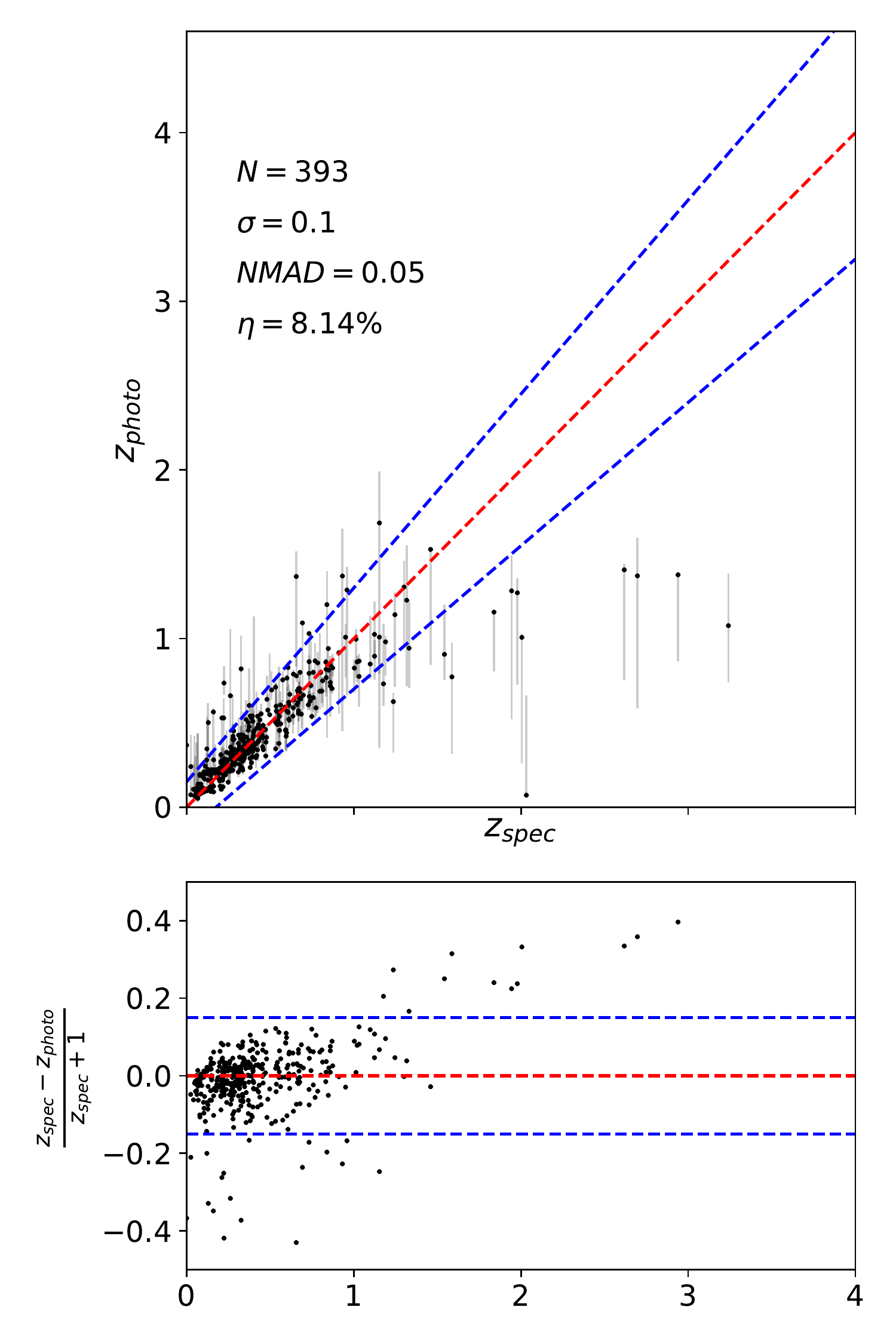}
	\end{subfigure} \hfill
	\begin{subfigure}[b]{0.32\textwidth}
		\caption{} \vspace{\adjheight}
		\includegraphics[width = \textwidth]{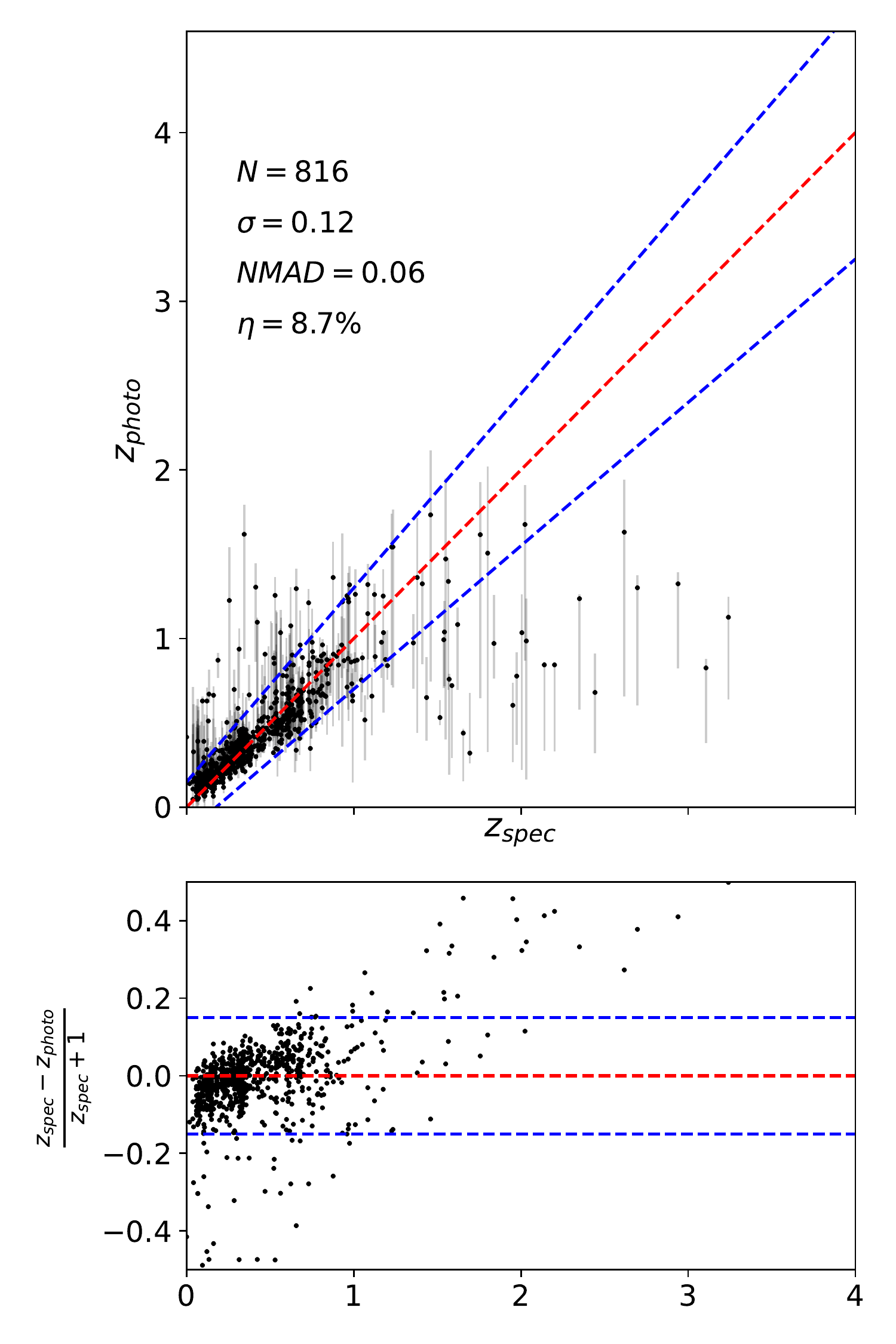}
	\end{subfigure} \hfill
	\begin{subfigure}[b]{0.32\textwidth}
		\caption{} \vspace{\adjheight}
		\includegraphics[width = \textwidth]{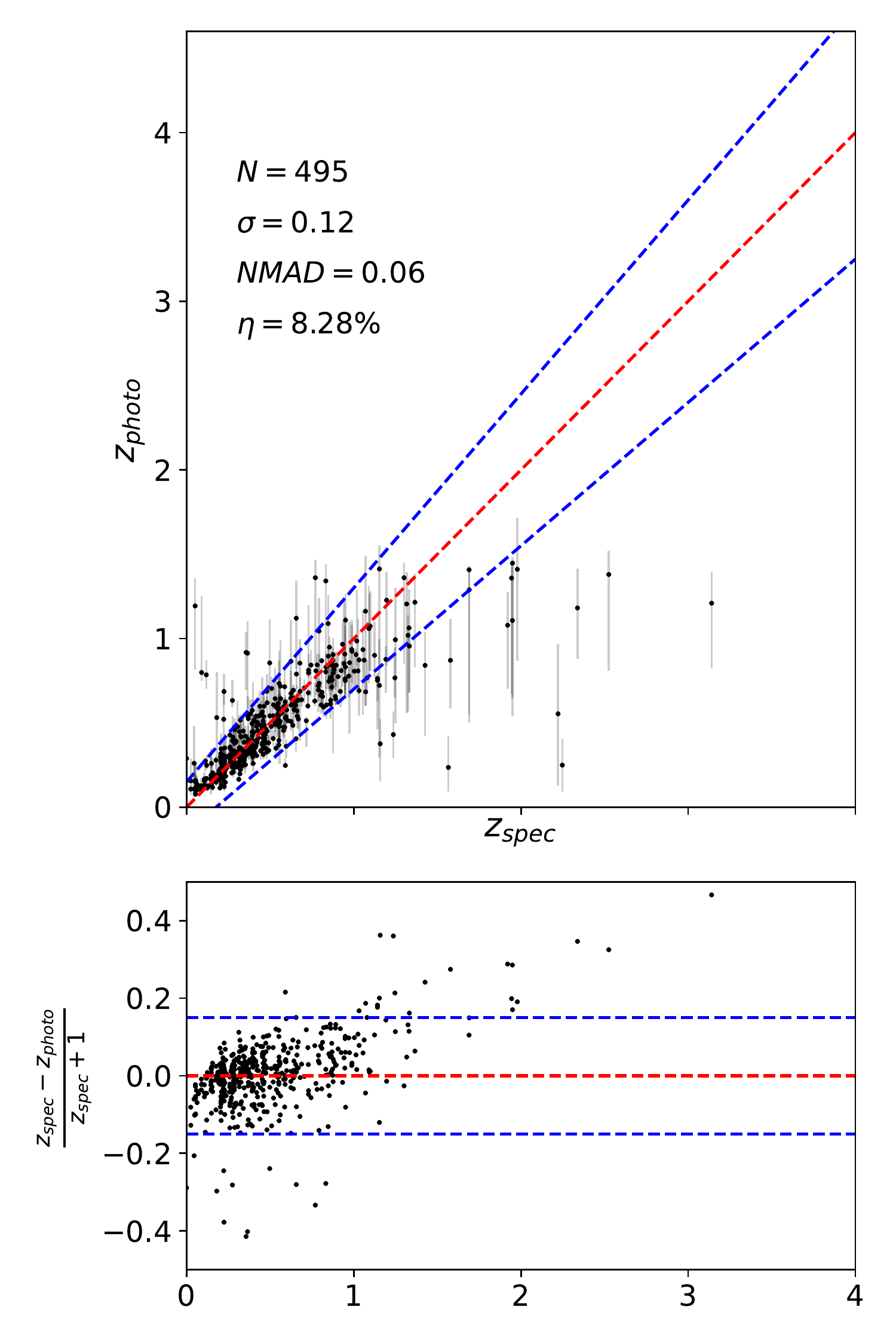}
	\end{subfigure} \hfill
\caption{Figures showing the results using the \ac{kNN} Regression algorithm using Euclidean Distance as the Distance metric, varying the data used for training (with the numbers corresponding to the Test ID in Table~\ref{table:RegressionTestID}). (REu1 -- Left) uses a random training sample, (REu2 -- centre) uses the \ac{ELAIS-S1} field as the training set, and (REu3 -- right) uses the \ac{eCDFS} field as the training set. The x-axes shows the measured spectroscopic redshift. The top-panel y-axes show the predicted redshift using the given model. The bottom-panel y-axes show the normalised residuals. The red-dashed line shows a perfect 1:1 prediction, and the blue-dashed lines show the decision boundaries based on the $\eta_{0.15}$ outlier rates.
}
\label{fig:results_euc_reg}
\vspace{1cm}
\end{figure*}

\begin{figure*}[ht!]
    \centering
	\begin{subfigure}[b]{0.32\textwidth}
		\caption{} \vspace{\adjheight}
		\includegraphics[width = \textwidth]{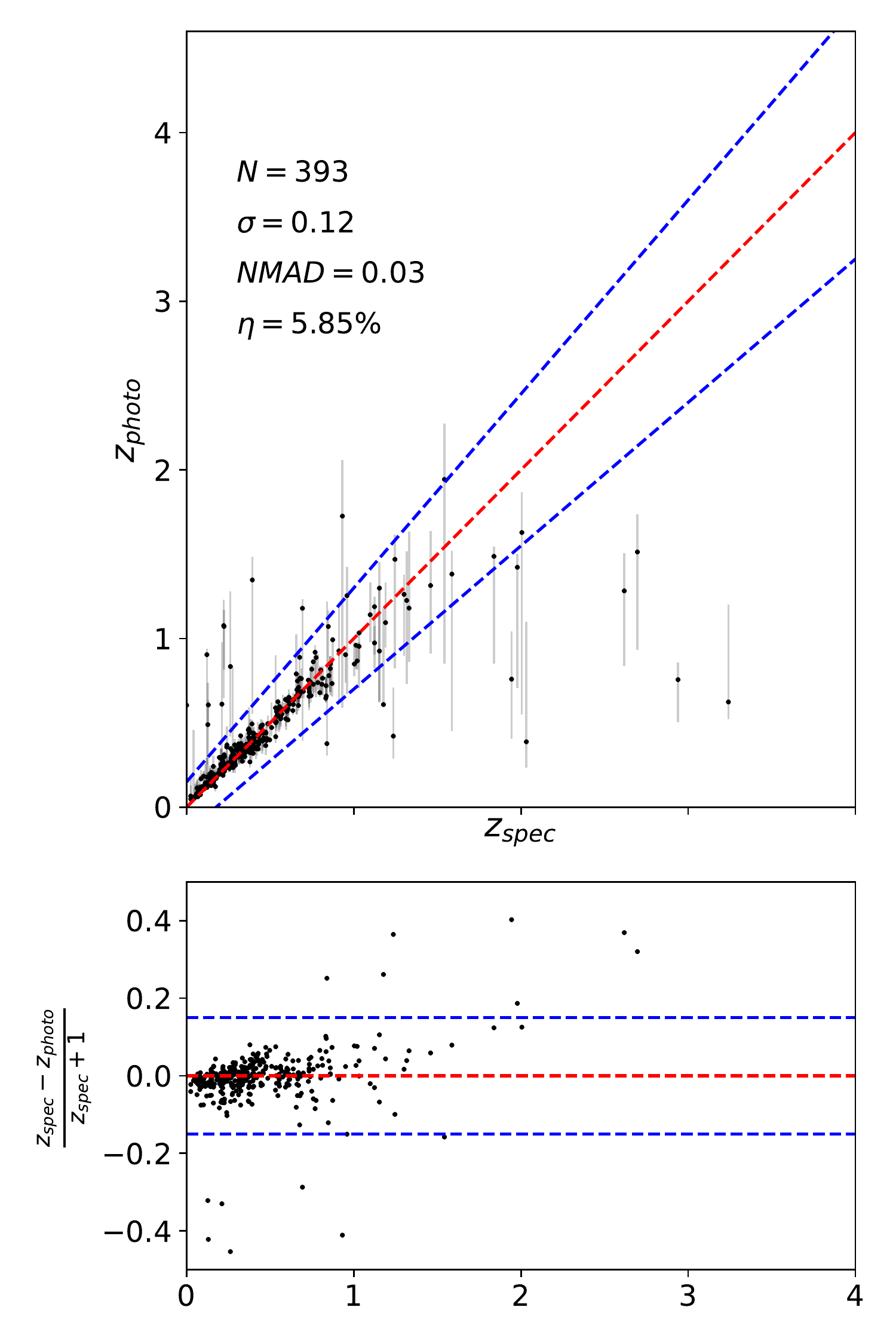}
	\end{subfigure} \hfill
	\begin{subfigure}[b]{0.32\textwidth}
		\caption{} \vspace{\adjheight}
		\includegraphics[width = \textwidth]{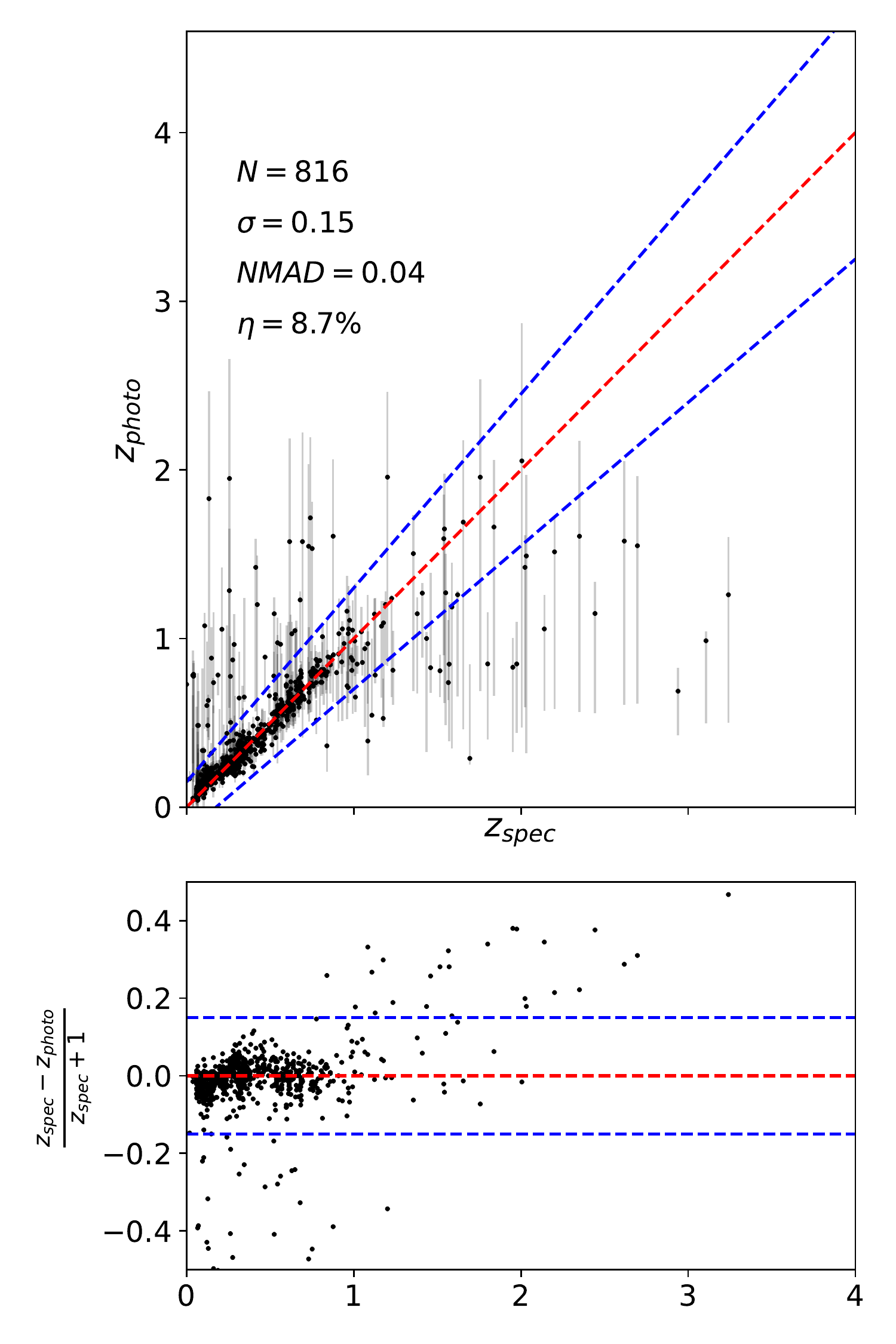}
	\end{subfigure} \hfill
	\begin{subfigure}[b]{0.32\textwidth}
		\caption{} \vspace{\adjheight}
		\includegraphics[width = \textwidth]{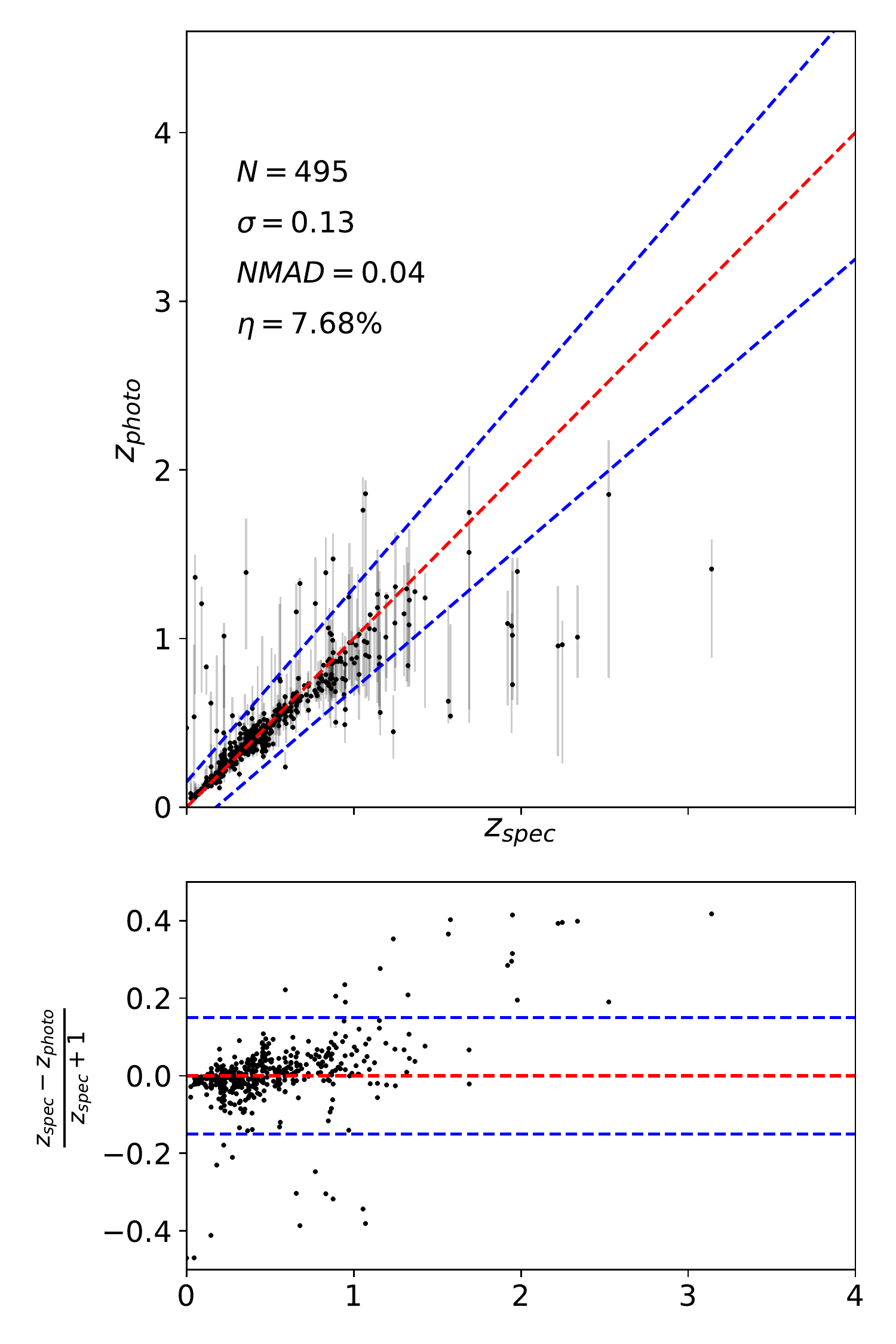}
	\end{subfigure} \hfill
\caption{Figures showing the results using the \ac{kNN} Regression algorithm using Mahalanobis Distance as the Distance metric, varying the data used for training (with the numbers corresponding to the Test ID in Table~\ref{table:RegressionTestID}). (RMa1 -- Left) uses a random training sample, (RMa2 -- centre) uses the \ac{ELAIS-S1} field as the training set, and (RMa3 -- right) uses the \ac{eCDFS} field as the training set. The x-axes shows the measured spectroscopic redshift. The top-panel y-axes show the predicted redshift using the given model. The bottom-panel y-axes show the normalised residuals. The red-dashed line shows a perfect 1:1 prediction, and the blue-dashed lines show the decision boundaries based on the $\eta_{0.15}$ outlier rates }
\label{fig:results_mahal_reg}
\vspace{1cm}
\end{figure*}

\begin{figure*}[ht!]
    \centering
	\begin{subfigure}[b]{0.32\textwidth}
		\caption{} \vspace{\adjheight}
		\includegraphics[width = \textwidth]{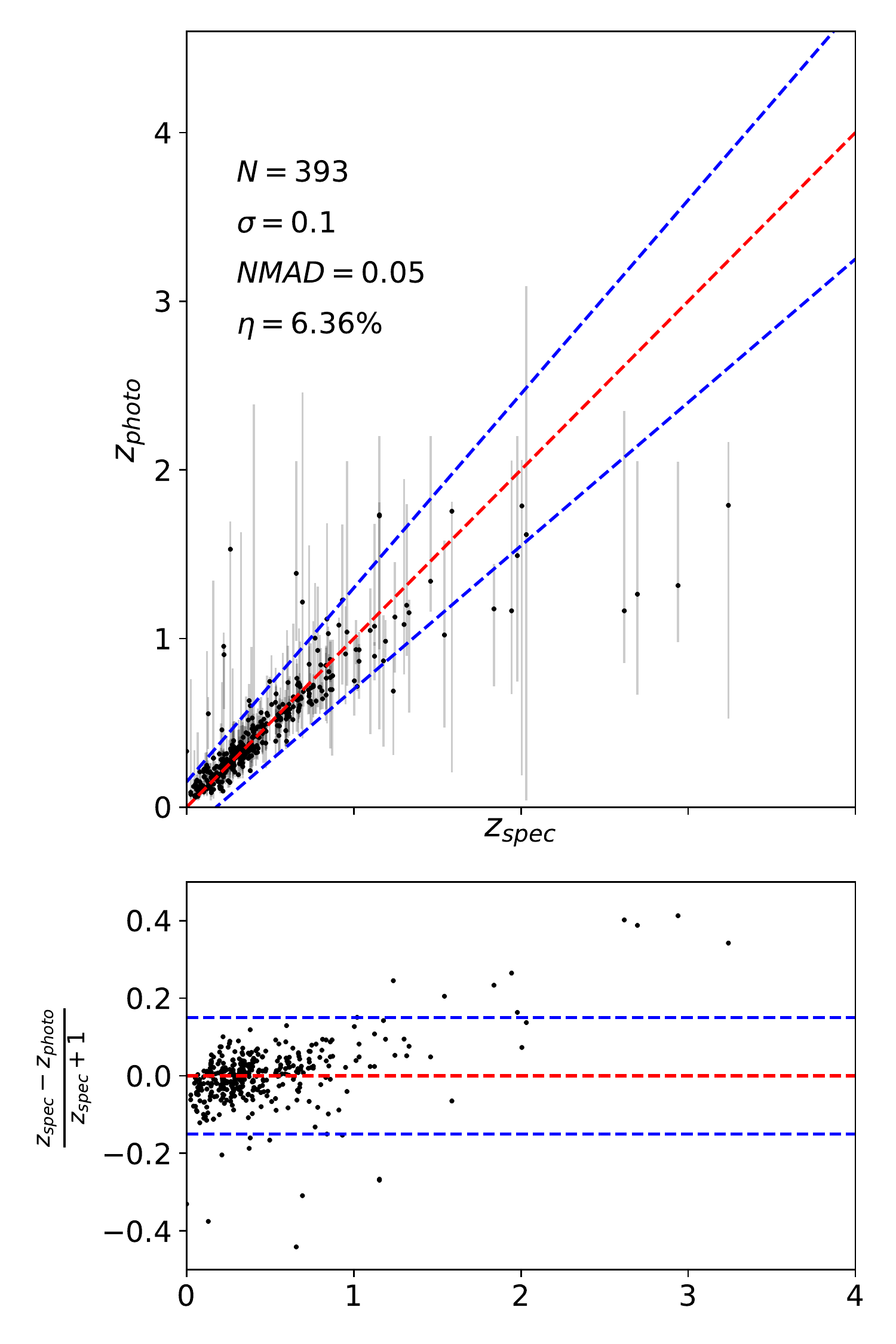}
	\end{subfigure} \hfill
	\begin{subfigure}[b]{0.32\textwidth}
		\caption{} \vspace{\adjheight}
		\includegraphics[width = \textwidth]{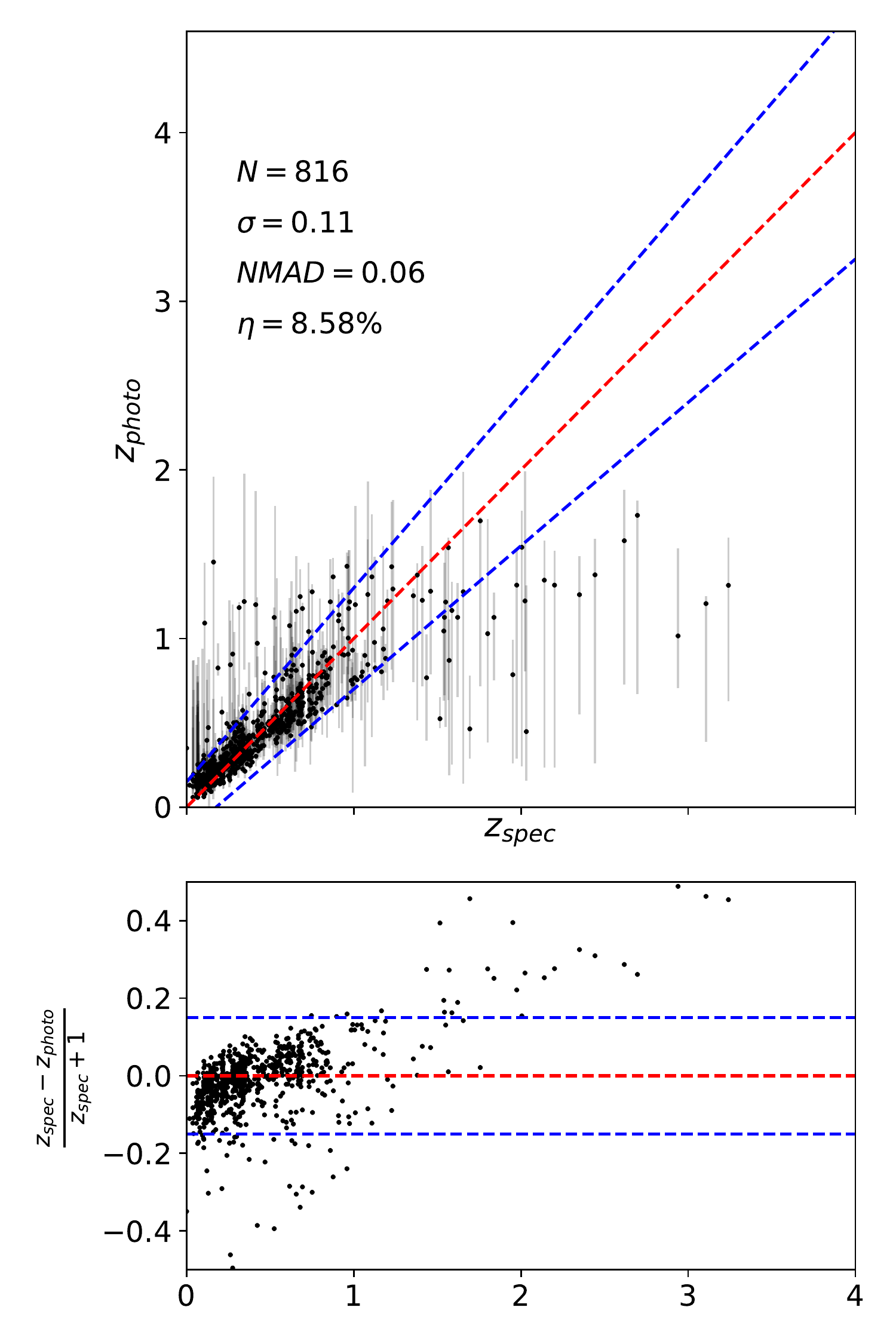}
	\end{subfigure} \hfill
	\begin{subfigure}[b]{0.32\textwidth}
		\caption{} \vspace{\adjheight}
		\includegraphics[width = \textwidth]{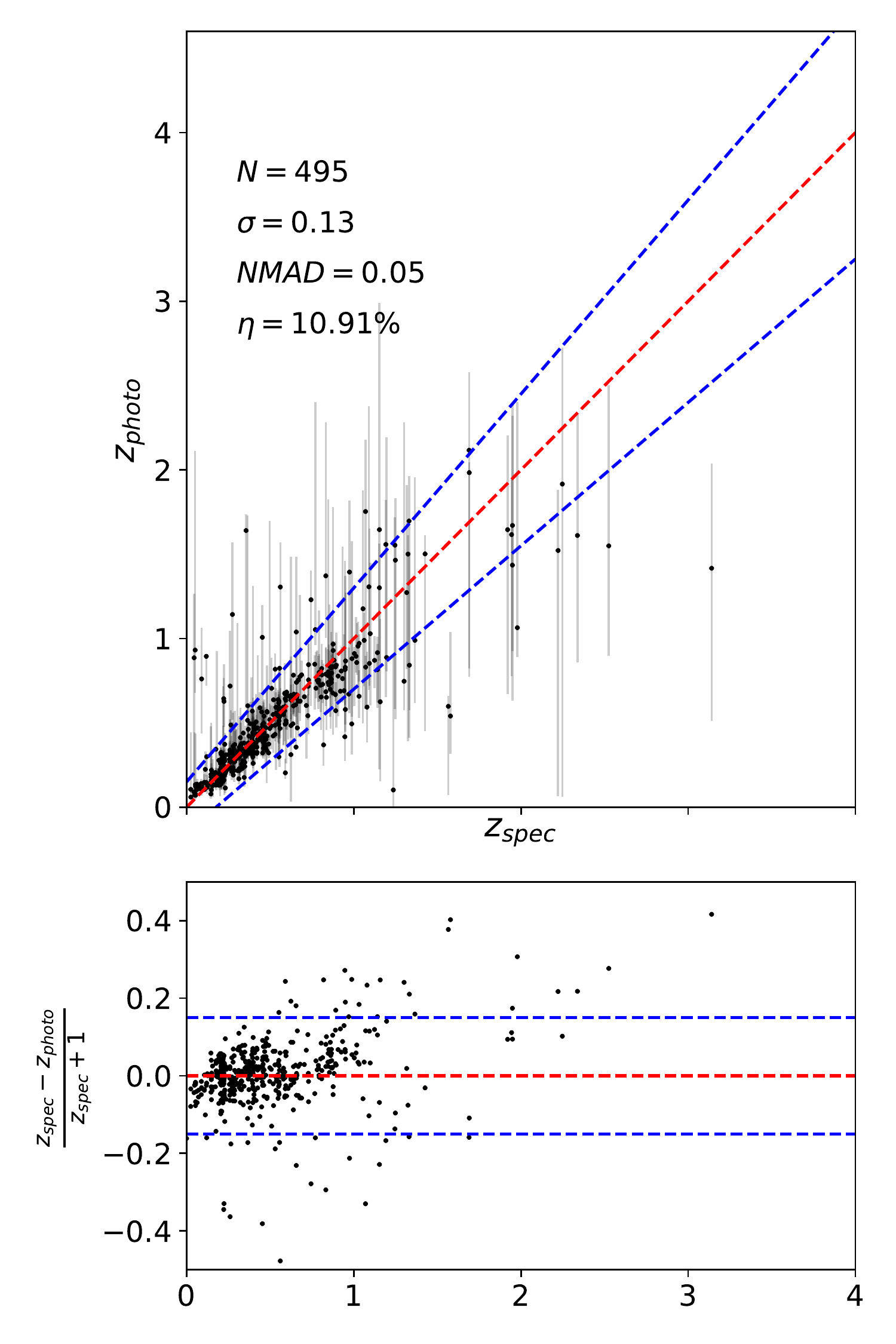}
	\end{subfigure} \hfill
\caption{Figures showing the results using the \ac{kNN} Regression algorithm using \ac{MLKR} Distance as the Distance metric, varying the data used for training (with the numbers corresponding to the Test ID in Table~\ref{table:RegressionTestID}). (RML1 -- Left) uses a random training sample, (RML2 -- centre) uses the \ac{ELAIS-S1} field as the training set, and (RML3 -- right) uses the \ac{eCDFS} field as the training set. The x-axes shows the measured spectroscopic redshift. The top-panel y-axes show the predicted redshift using the given model. The bottom-panel y-axes show the normalised residuals. The red-dashed line shows a perfect 1:1 prediction, and the blue-dashed lines show the decision boundaries based on the $\eta_{0.15}$ outlier rates.}
\label{fig:results_mlkr_reg}
\end{figure*}

\begin{figure*}[ht!]
    \centering
	\begin{subfigure}[b]{0.32\textwidth}
		\caption{} \vspace{\adjheight}
		\includegraphics[width = \textwidth]{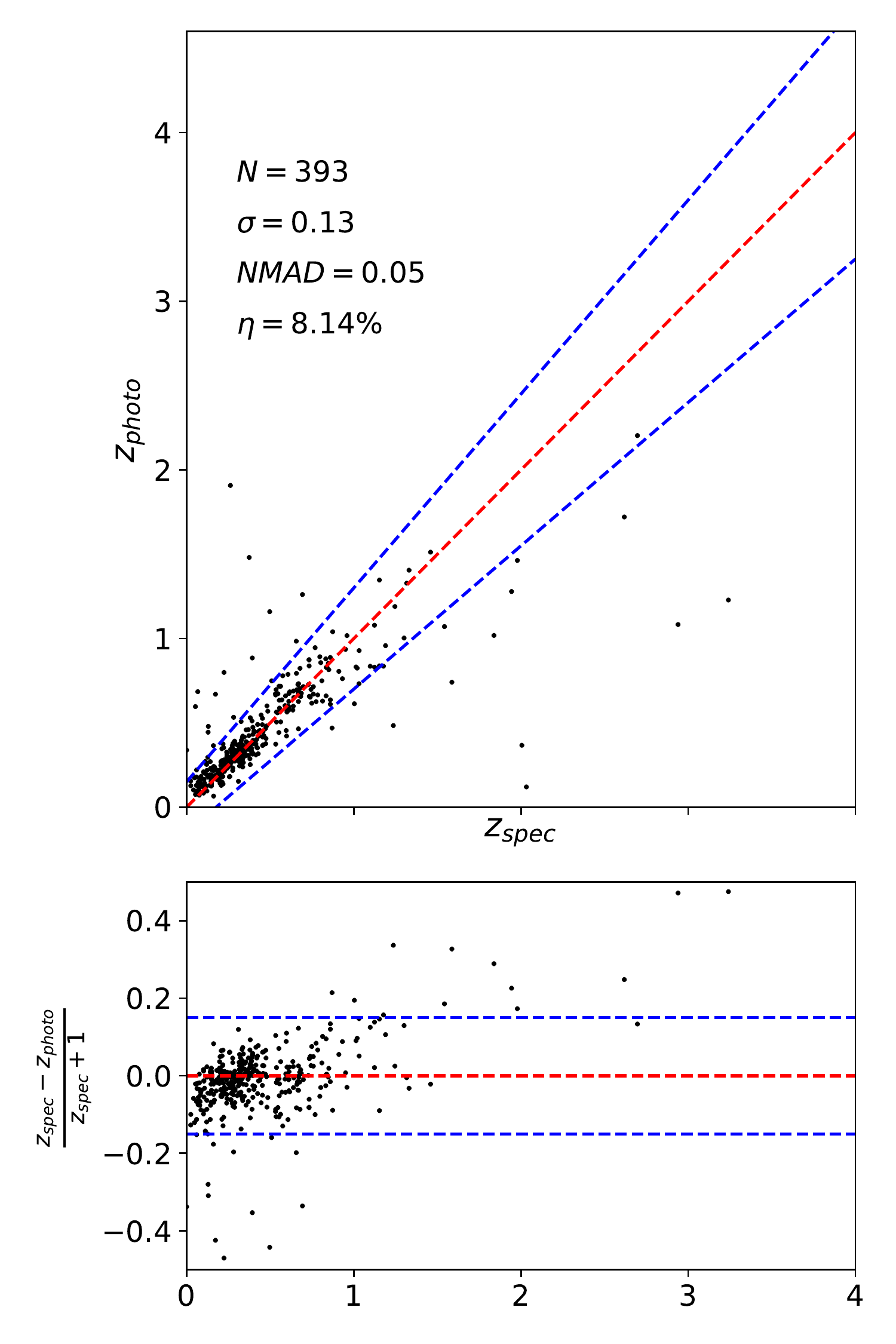}
	\end{subfigure} \hfill
	\begin{subfigure}[b]{0.32\textwidth}
		\caption{} \vspace{\adjheight}
		\includegraphics[width = \textwidth]{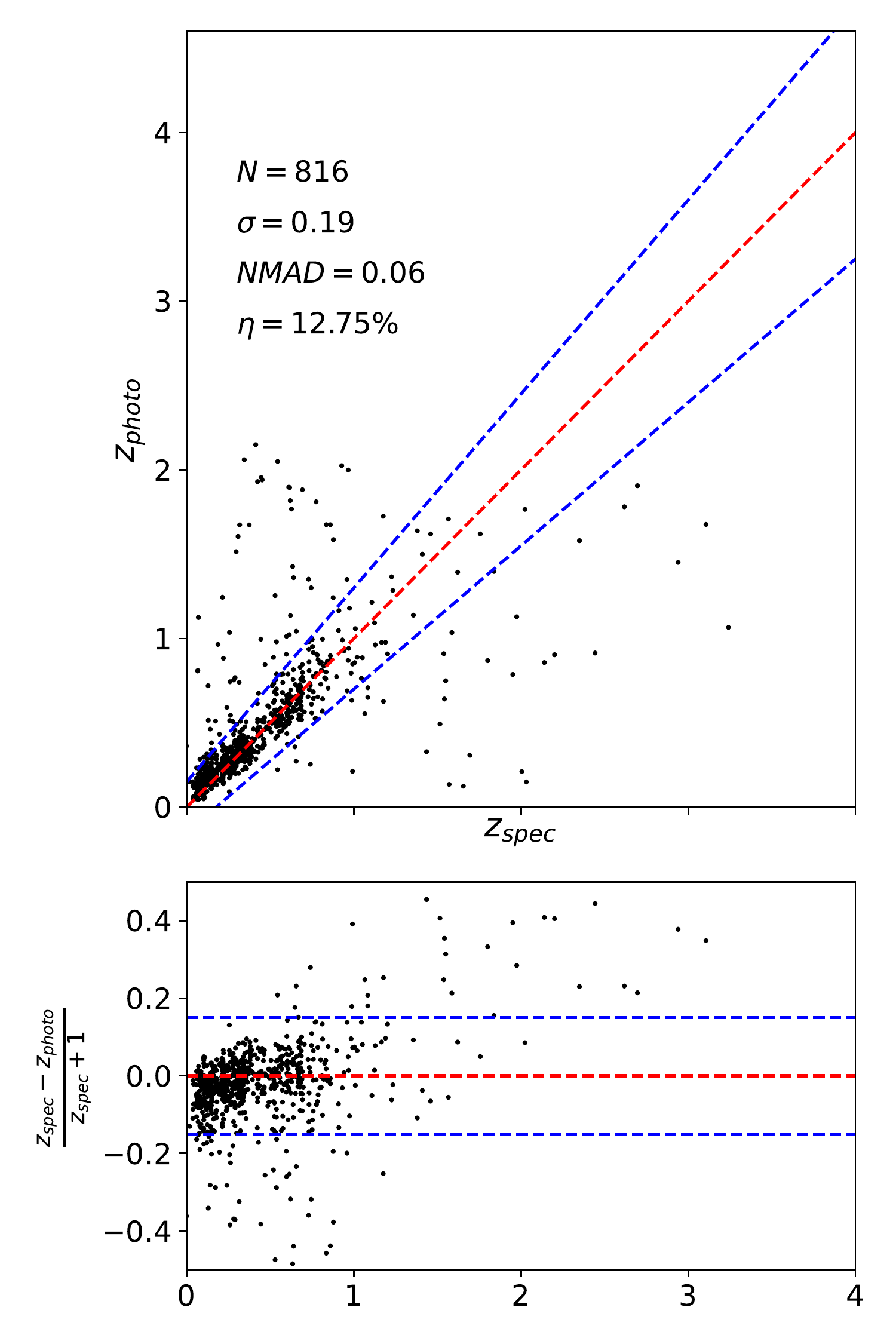}
	\end{subfigure} \hfill
	\begin{subfigure}[b]{0.32\textwidth}
		\caption{} \vspace{\adjheight}
		\includegraphics[width = \textwidth]{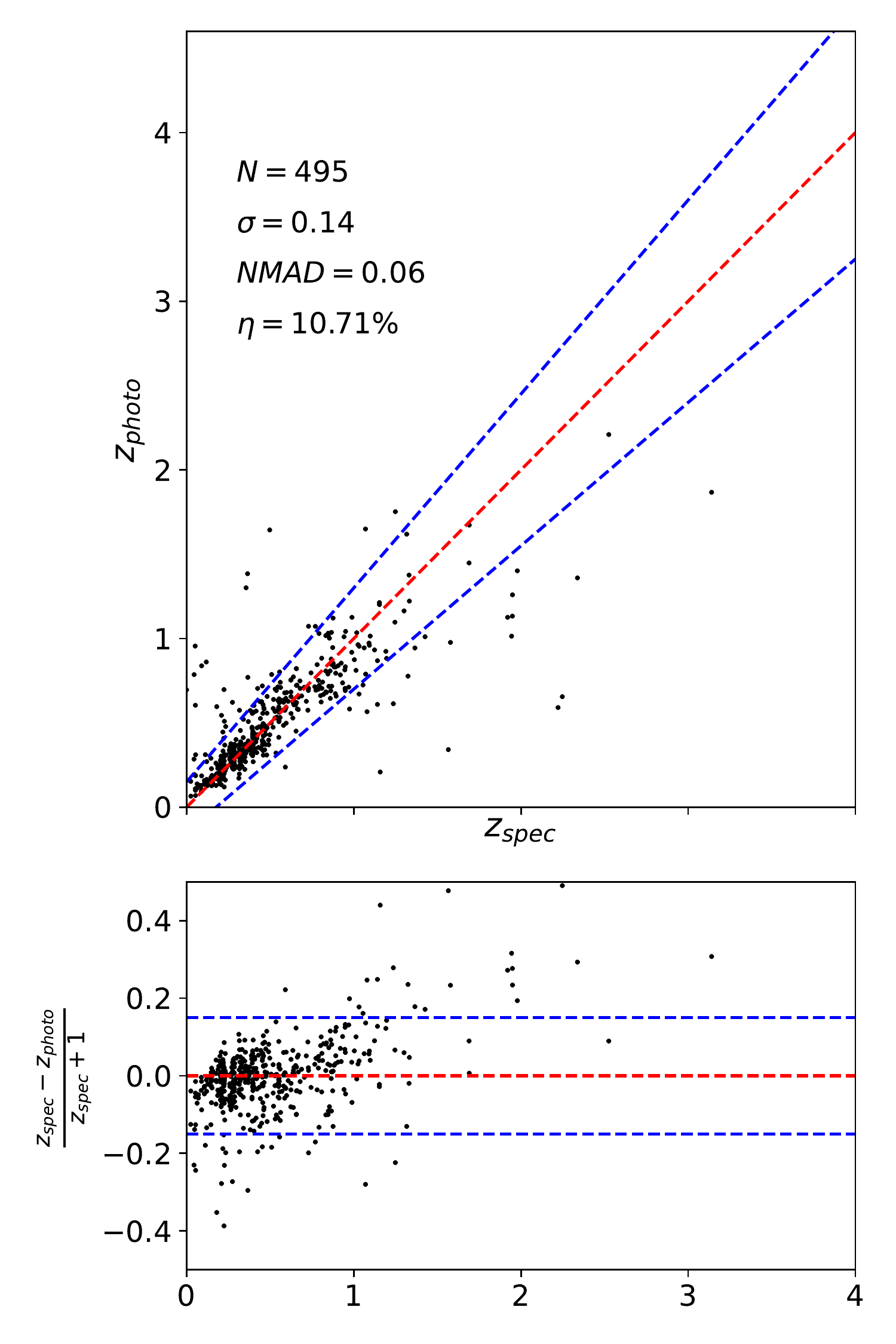}
	\end{subfigure} \hfill
\caption{Figures showing the results using the \ac{RF} Regression algorithm, varying the data used for training (with the numbers corresponding to the Test ID in Table~\ref{table:RegressionTestID}). (RRf1 -- Left) uses a random training sample, (RRf2 -- centre) uses the \ac{ELAIS-S1} field as the training set, and (RRf3 -- right) uses the \ac{eCDFS} field as the training set. The x-axes shows the measured spectroscopic redshift. The top-panel y-axes show the predicted redshift using the given model. The bottom-panel y-axes show the normalised residuals. The red-dashed line shows a perfect 1:1 prediction, and the blue-dashed lines show the decision boundaries based on the $\eta_{0.15}$ outlier rates.}
\label{fig:results_rf_reg}
\end{figure*}

Based on the Regression tests outlined in Table~\ref{table:RegressionTestID}, we present the results in a series of plots (Figures~\ref{fig:results_euc_reg}, \ref{fig:results_mahal_reg}, \ref{fig:results_mlkr_reg}, and \ref{fig:results_rf_reg}). 

In each plot three subfigures are presented, with Subfigure A representing the random training set, Subfigure B representing the training set built using galaxies from the \ac{ELAIS-S1} field, and Subfigure C representing the training set built using galaxies from the \ac{eCDFS} field. The subfigures are split into two plots -- the top plot showing the measured spectroscopic redshift (x-axis) compared with the predicted redshift (y-axis), with a perfect 1:1 correlation (red dashed line) and the $\eta_{0.15}$ outlier rate boundaries (blue dashed lines) shown. The bottom plot shows the measured spectroscopic redshift (x-axis) against the residuals, again with the perfect 1:1 correlation (red dashed line) and the $\eta_{0.15}$ outlier rate boundaries (blue dashed lines) shown. 

These results are summarised in Table~\ref{table:regressionResults}.

\begin{table}[]
    \caption{Results using the regression based algorithms. The Test column relates to the tests defined in Table~\ref{table:RegressionTestID}. The best result for each metric is highlighted in bold.}
    \label{table:regressionResults}
    \resizebox{0.5\textwidth}{!}{
        \begin{tabular}{cccccccc}
        \toprule
            Test & $k$ / & $\eta_{0.15}$ & $\eta_{2\sigma}$ & $R^2$ & \ac{MSE} & $\sigma$ & $\sigma_\mathrm{NMAD}$ \\
                 & Trees & (\%)          & (\%)             & Value                &                           &          &                            \\
                 \midrule
            REu1   & 5           & 8.14          & 6.61             & 0.65                 & 0.06                      & \textbf{0.10}      & 0.05                       \\
            REu2   & 5           & 8.70          & 5.51             & 0.60                 & 0.06                      & 0.12     & 0.06                       \\\smallskip
            REu3   & 11          & 8.28           & 4.85             & 0.53                 & 0.09                       & 0.12     & 0.06                       \\
            RMa1   & 5           & \textbf{5.85}          & 5.09             & 0.59                 & 0.07                     & 0.12     & \textbf{0.03}                       \\
            RMa2   & 3           &  8.70         & 5.39             & 0.58                 & 0.07                      & 0.15     & 0.04                       \\\smallskip
            RMa3   & 5          & 7.68          & 5.45             & 0.54                 & 0.09                      & 0.13     & 0.04                       \\
            RML1   & 4           & 6.36          & \textbf{4.07}             & \textbf{0.73}                 & \textbf{0.05}                      & \textbf{0.10}      & 0.05                       \\
            RML2   & 9           & 8.58          & 4.78             & 0.68                 & \textbf{0.05}                      & 0.11     & 0.06                       \\\smallskip
            RML3   & 5           & 10.91         & 4.44             & 0.58                 & 0.08                      & 0.13     & 0.05                      \\
            RRf1  & 44          & 8.14          & 4.84             & 0.59                 & 0.07                      & 0.13     & 0.05                       \\
            RRf2  & 21          & 12.75         & 5.88             & 0.32                 & 0.11                      & 0.19      & 0.06                       \\
            RRf3  & 23          & 10.71         & 4.65             & 0.54                 & 0.09                      & 0.14     & 0.06                       \\
            \bottomrule
        \end{tabular}
    }
\end{table}

\renewcommand{\thesubfigure}{REu\arabic{subfigure})}

\renewcommand{\thesubfigure}{RMa\arabic{subfigure})}

\renewcommand{\thesubfigure}{RML\arabic{subfigure})}

\renewcommand{\thesubfigure}{RRf\arabic{subfigure})}

We show that the lowest $\eta_{0.15}$ outlier rate is achieved using the \ac{kNN} algorithm paired with the Mahalanobis distance metric, and is statistically different from most other algorithms (\ac{kNN} using Euclidean Distance: p value = 0.0183, and the \ac{RF} algorithm: p value = 0.0183 compared with \ac{kNN} using the Mahalanobis distance metric as a baseline). The \ac{kNN} algorithm using the \ac{MLKR} distance metric (a Mahalanobis-like distance metric) is not statistically significantly different (p value = 0.5750). However, all results (including the \ac{RF} algorithm) suffer from the same issues of under-predicting high redshift ($z > 1$) galaxies. 

The randomly selected training sets typically achieve lower $\eta_{0.15}$ outlier rates than the training sets built using one field only, however, neither field-based training set provides consistently better $\eta_{0.15}$ outlier rates than the other. This is confirmed statistically, with no statistically significant result measured (p value = 0.2072).

\subsection{Classification}

\begin{figure}
    \centering
    \includegraphics[width = 0.49\textwidth]{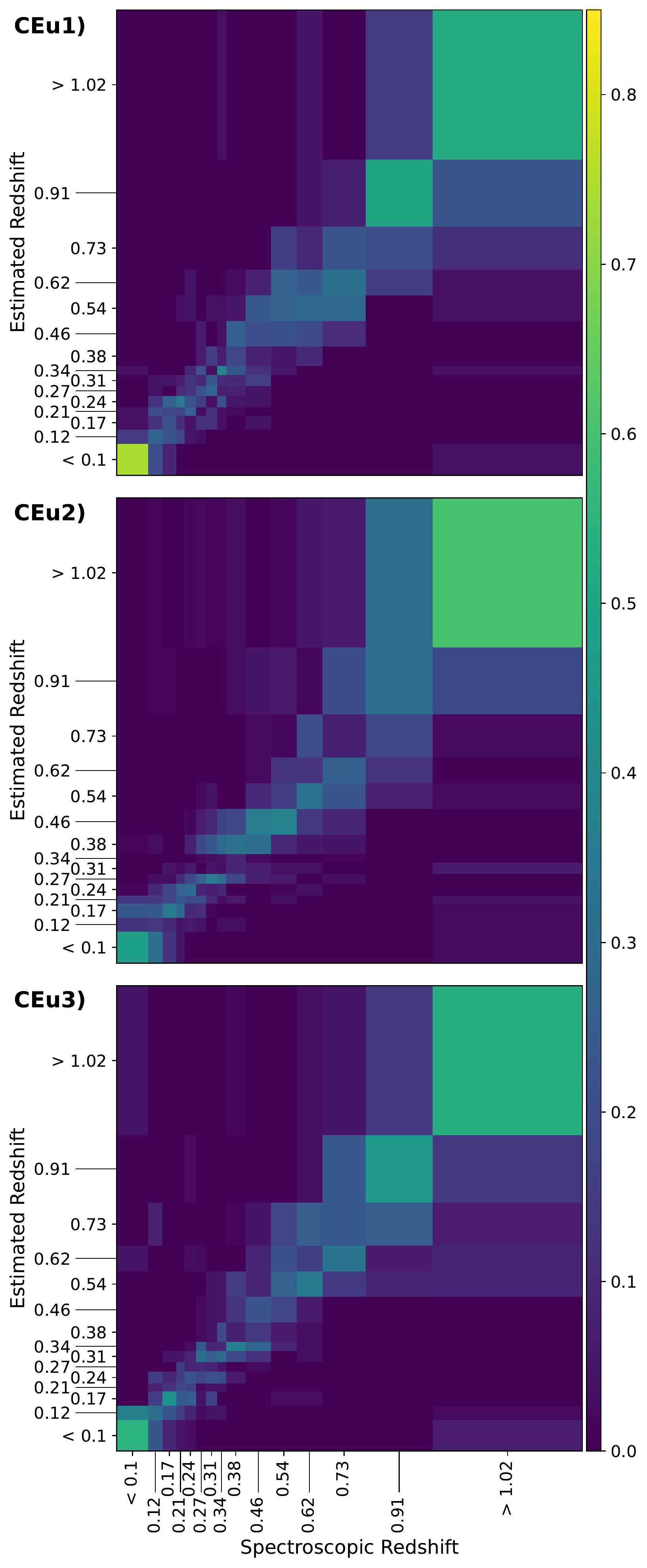}
    \caption{The confusion matrix of the \ac{kNN} classification tests using the Euclidean distance metric. The x-axis of each plot shows the measured spectroscopic redshift, and the y-axis shows the estimated redshift. The size of the boxes is scaled based on the width of the bin used for classification. (CEu1 -- Top) uses a random training sample, (CEu2 -- Middle) uses a training sample consisting of the \ac{ELAIS-S1} field, and (CEu3 -- Bottom) uses a training sample consisting of the \ac{eCDFS} field.}
    \label{fig:results_knn_euclid_class}
\end{figure}

\begin{figure}
    \centering
    \includegraphics[width = 0.49\textwidth]{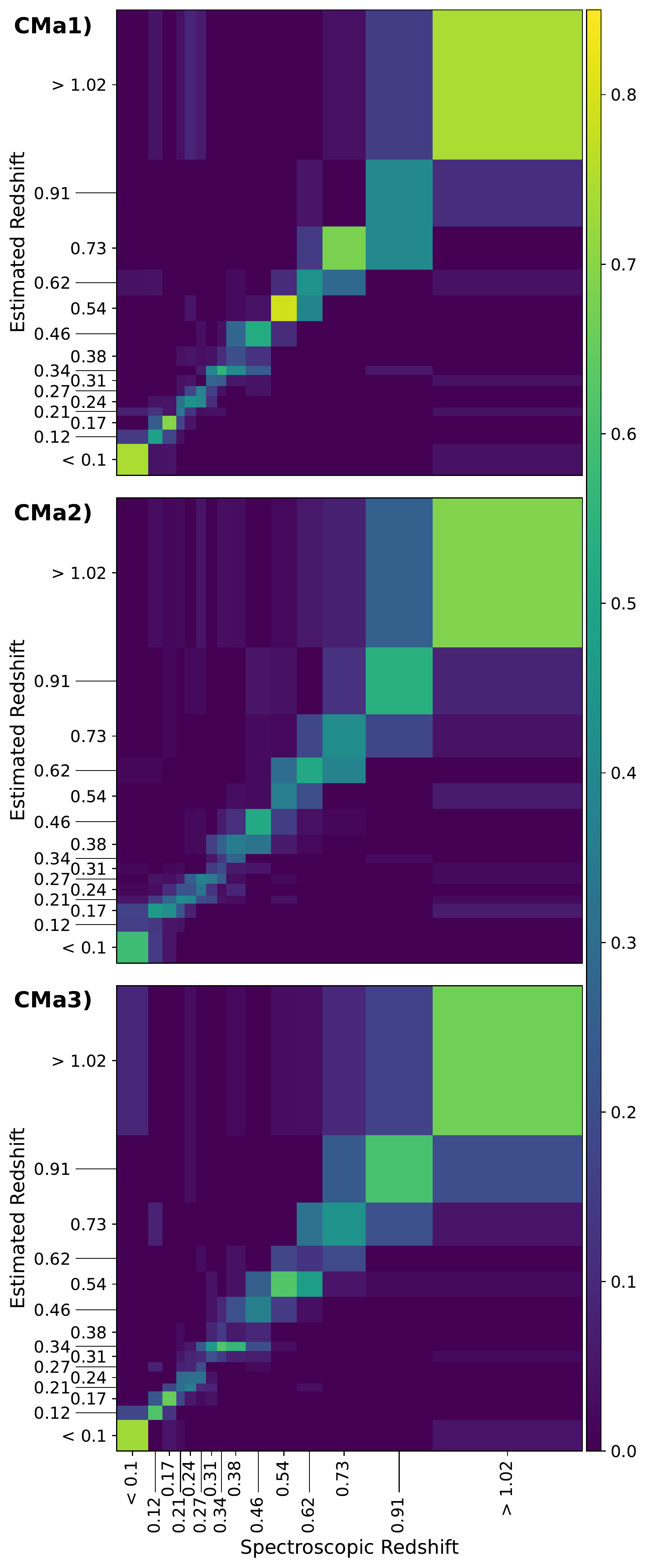}
    \caption{The confusion matrix of the \ac{kNN} classification tests using the Mahalanobis distance metric. The x-axis of each plot shows the measured spectroscopic redshift, and the y-axis shows the estimated redshift. The size of the boxes is scaled based on the width of the bin used for classification. (CMa1 -- Top) uses a random training sample, (CMa2 -- Middle) uses a training sample consisting of the \ac{ELAIS-S1} field, and (CMa3 -- Bottom) uses a training sample consisting of the \ac{eCDFS} field. }
    \label{fig:results_knn_mahal_class}
\end{figure}

\begin{figure}
    \centering
    \includegraphics[width = 0.49\textwidth]{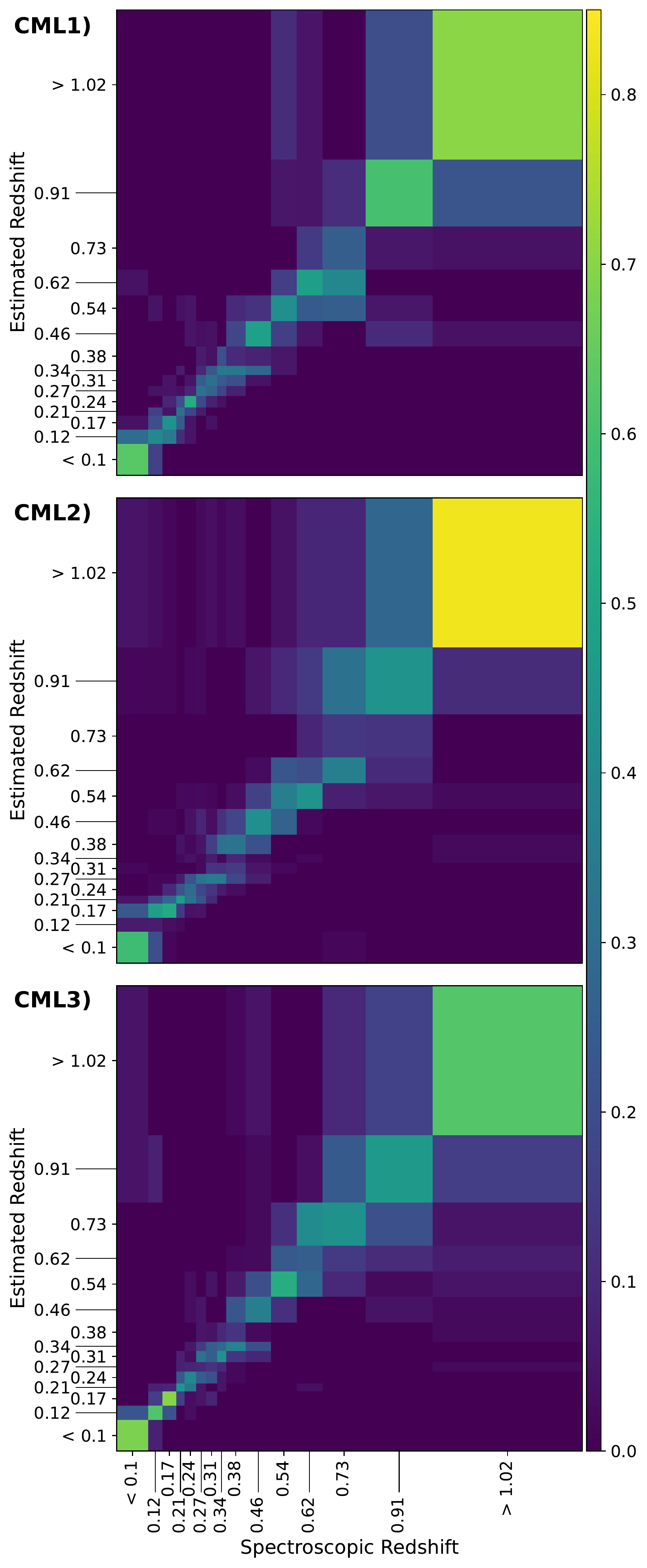}
    \caption{The confusion matrix of the \ac{kNN} classification tests using the \ac{LMNN} learned distance metric. The x-axis of each plot shows the measured spectroscopic redshift, and the y-axis shows the estimated redshift. The size of the boxes is scaled based on the width of the bin used for classification. (CML1 -- Top) uses a random training sample, (CML2 -- Middle) uses a training sample consisting of the \ac{ELAIS-S1} field, and (CML3 -- Left) uses a training sample consisting of the \ac{eCDFS} field.}
    \label{fig:results_knn_lmnn_class}
\end{figure}

\begin{figure}
    \centering
    \includegraphics[width = 0.49\textwidth]{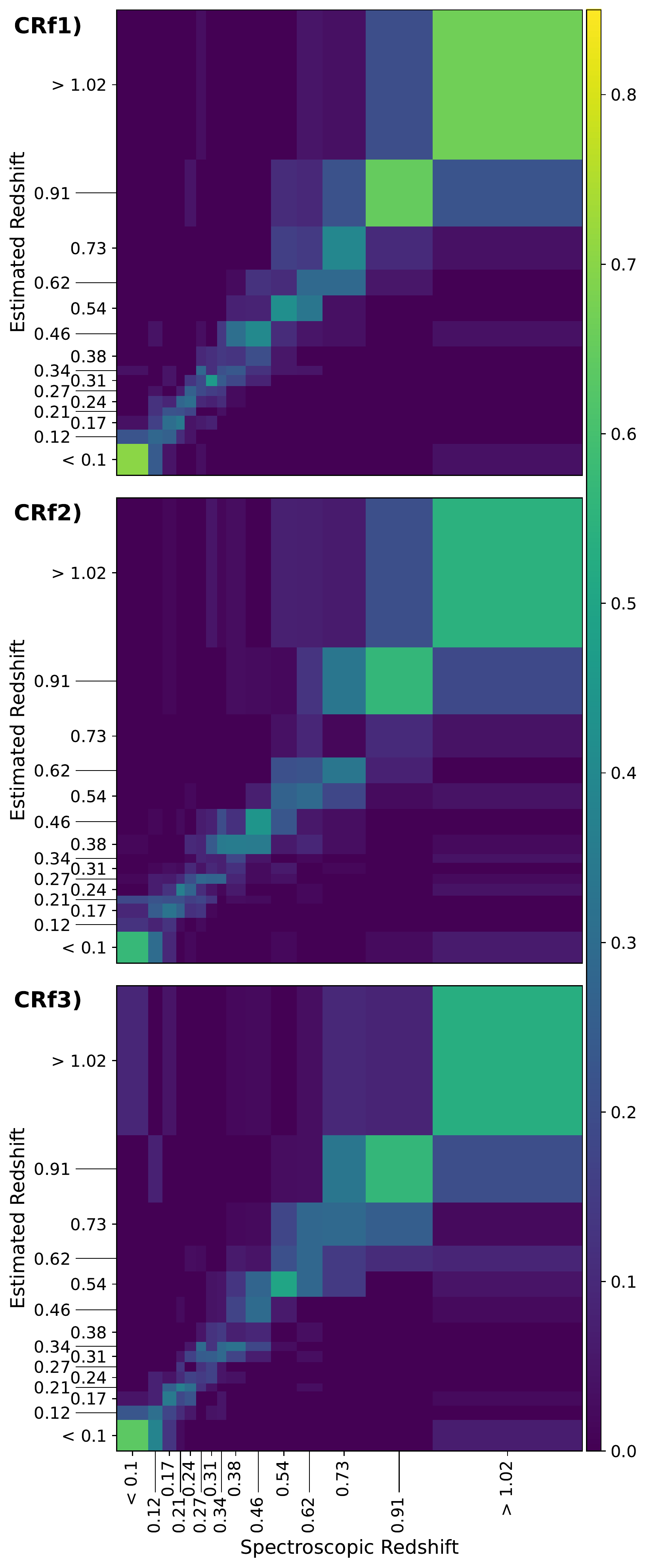}
    \caption{The confusion matrix of the \ac{RF} classification tests. The x-axis of each plot shows the measured spectroscopic redshift, and the y-axis shows the estimated redshift. The size of the boxes is scaled based on the width of the bin used for classification. (CRf1 -- Top) uses a random training sample, (CRf2 -- Middle) uses a training sample consisting of the \ac{ELAIS-S1} field, and (CRf3 -- Bottom) uses a training sample consisting of the \ac{eCDFS} field. }
    \label{fig:results_rf_class}
\end{figure}

Based on the Classification tests outlined in Table~\ref{table:ClassificationTestID}, we show the classification-based results for the \ac{kNN} algorithm -- using Euclidean distance (Figure~\ref{fig:results_knn_euclid_class}), Mahalanobis distance (Figure~\ref{fig:results_knn_mahal_class}) and the \ac{LMNN} learned distance metric (Figure~\ref{fig:results_knn_lmnn_class}) -- and the \ac{RF} algorithm (Figure~\ref{fig:results_rf_class}), summarised in Table~\ref{table:classification_results}.

\begin{figure}
    \centering
    \includegraphics[width = 0.32\textwidth]{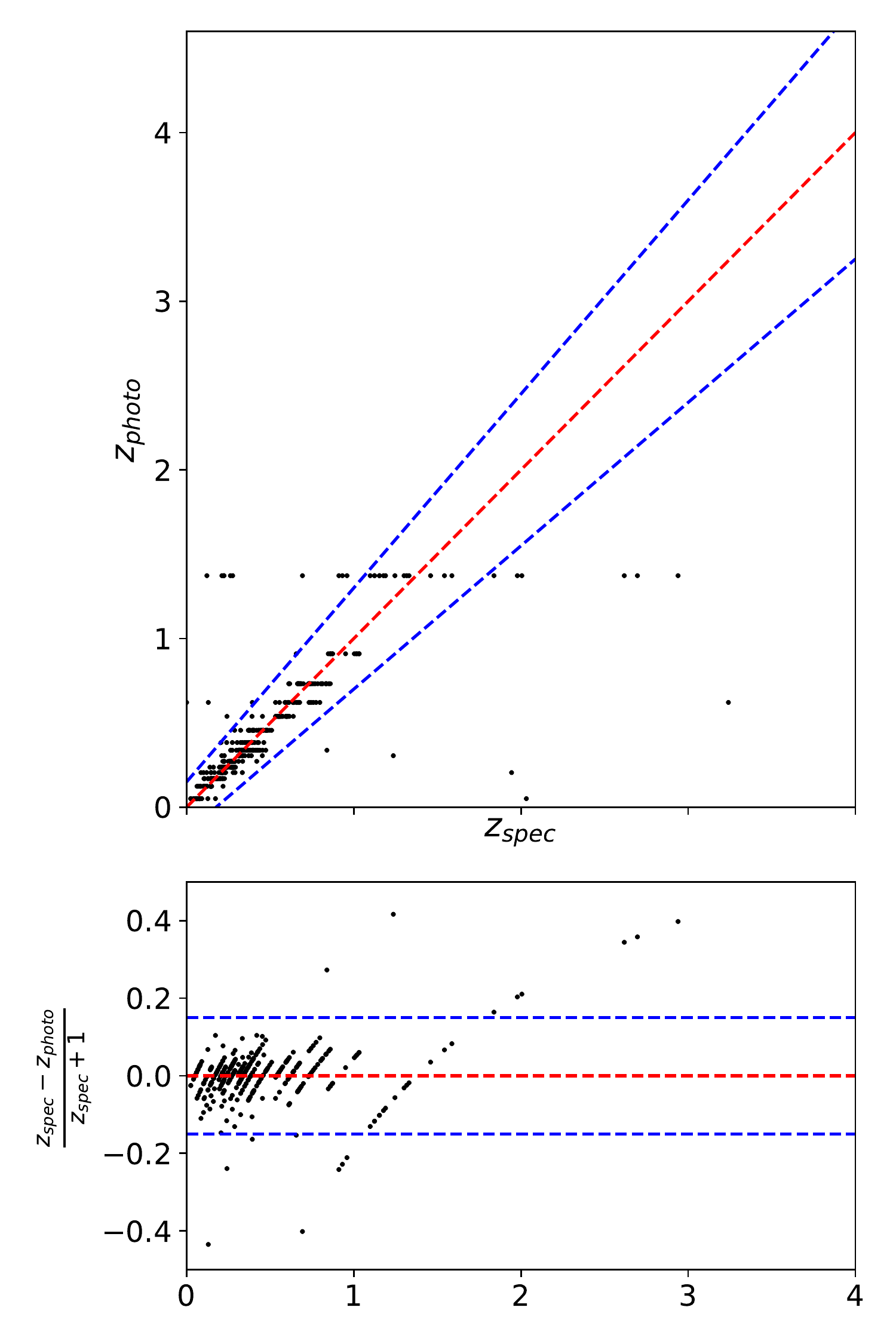}
    \caption{An example scatter plot similar to Figure~\ref{fig:results_euc_reg} for the CMa1 test, showing the effect of binning and classifying. }
    \label{fig:results_class_scatter}
\end{figure}

We present our classification-based results using scaled confusion matrices, where the x-axis shows the measured spectroscopic redshift, the y-axis shows the predicted redshift, the colour shows the density of the objects in that chosen bin, and the width of each bin is proportional to the range of redshift values represented by the bin. An example scatter plot (in the same style as Figure~\ref{fig:results_euc_reg}), demonstrating the effect of binning on the classification results when compared against the original spectroscopic measurements is shown in Figure~\ref{fig:results_class_scatter}.

\begin{table}[]
    \caption{Results using the classification based algorithms. The Test column relates to the tests defined in Table~\ref{table:ClassificationTestID}. The best result for each metric is highlighted in bold}
    \label{table:classification_results}
        \begin{tabular}{llllllllll}
                 \toprule
        Test & $k$ / & $\eta_{0.15}$ & $\eta_{2\sigma}$ & Acc  & $\sigma$ \\
             & Trees & (\%)          & (\%)             &      &          \\
                 \midrule
        CEu1   & 6     & 8.91          & 4.07             & 0.30 & 0.10     \\
        CEu2   & 5     & 12.13         & 3.55             & 0.24 & 0.14     \\\smallskip
        CEu3   & 8     & 11.31         & 5.05             & 0.27 & 0.13     \\
        CMa1   & 11     & \textbf{5.85}          & 3.57   & \textbf{0.50} & 0.14     \\
        CMa2   & 6     & 7.60          & 4.41             & 0.37 & 0.14     \\\smallskip
        CMa3   & 13    & 7.88          & \textbf{3.43}    & 0.40 & 0.13     \\
        CML1   & 7     & 6.36          & 5.85             & 0.40 & \textbf{0.09}     \\
        CML2   & 8    & 9.93          & 4.04             & 0.32 & 0.16     \\\smallskip
        CML3   & 7     & 9.90          & 4.04             & 0.40 & 0.13     \\
        CRf1  & 53    & 7.12          & 3.82             & 0.36 & 0.10     \\
        CRf2  & 39    & 10.54         & 4.41             & 0.27 & 0.14     \\
        CRf3  & 32    & 10.91         & 4.24             & 0.33 & 0.15     \\
                 \bottomrule
        \end{tabular}
\end{table}

The \ac{kNN} algorithm paired with the Mahalanobis distance metric provides the lowest $\eta_{0.15}$ and $\eta_{2\sigma}$ outlier rates, as well as performing the best in terms of traditional \ac{ML} classification metrics (accuracy, precision, recall and F1 score). However, while the results using the Mahalanobis distance metric are statistically significantly better than those using the Euclidean distance metric (p\,=\,0.00619), they are not statistically different from the \ac{LMNN} learned distance metric (a Mahalanobis-like distance metric; p\,=\,0.4276), or the \ac{RF} algorithm (p\,=\,0.8913). Again, we note that for most algorithms, the highest redshift galaxies remain a problem for estimation, however, the results can change significantly based on the distance metric used -- for example, the $z > 1.02$ bin was correctly predicted $\sim$50\% of the time using the \ac{kNN} algorithm paired with Euclidean distance, however, using the Mahalanobis distance metric brought this up to $\sim$74\%.

As with the regression tests, the random training sample outperformed the training sets built from a single field in the primary $\eta_{0.15}$ error metric, however, the results are statistically insignificant (p\,=\,0.4397).

\section{Discussion}
Most previous work has treated the estimation of galaxy redshift as a regression task -- estimating a continuous value for redshift. We have shown that the kNN algorithm -- when using either a learned distance metric, or the Mahalanobis distance metric -- outperforms the RF algorithm, particularly at a redshift of $z < 1$. At a redshift of $z > 1$, both the kNN and RF have systematic under-estimations, likely caused by the unbalanced training sample used, with the vast majority of samples at lower redshifts.

Here we try a different approach, by treating the problem of calculating point estimates of redshifts as a classification problem. Previous works have calculated probability density functions (PDFs) for galaxies by finely binning data and treating it as a classification problem \citep{gerdes,pasquet-itam_deep_2018,eriksen}. However, this is subtly different to this work as the primary purpose here is to identify the high-redshift galaxies -- not to generate PDFs. We balance the data by binning the data into 15 bins with equal numbers of sources. By allowing this coarse mapping, we ensure that all redshift values we are attempting to predict are equally represented, and can therefore obtain better results at the higher redshift ranges. In particular, we have included a $z > 1$ bin, which in our best results (the kNN algorithm using the Mahalanobis distance metric) we predict in 74\% of cases (85\% of galaxies at $z > 0.8$ predicted in the highest two bins). Again, the kNN algorithm is able to outperform the RF algorithm.

In addition to the regression and classification tests, we tested using different training samples. The first training sample is made up of a random selection across both the ELAIS-S1 and eCDFS fields. The second and third samples were made up of only one of the fields, which was then used to predict the other. As expected, the random sample modelled the test sets better -- though not significantly -- than the training samples taken from a single field only. This is likely due to observational differences -- while the surveys used in this work were designed to be as homogeneous as possible, there will always be slight differences.

One such difference is the number of sources in each field. While there is not expected to be differences in the actual source counts between these fields, there is the potential issue of minor observational issues (for example, in the radio regime, there may have been more radio-frequency interference in one set of observations than another. In the optical regimes, it may have been that there was slightly different sky conditions affecting the observations) affecting the fields differently. 

\subsection{Comparison with Previous Works}
Fair comparisons with previous works are often difficult due to different features being selected --- often studies are conducted on well-observed, feature rich fields like the \ac{COSMOS} field, providing a wealth of UV, NIR and X-ray data that is typically not available for the majority of the sky --- and different selection methods --- selecting sources from the SDSS Galaxy survey, where the redshift range is restricted towards higher redshift and typically don't have a radio-counterpart, which are shown by \citet{norris_comparison_2018} to be a more difficult challenge. Where the data sets have been AGN selected, using similar features as in \citet[][ Lockman Hole AGN Experiments]{Duncan_2021}, we find our results are highly competitive, providing a lower outlier rate ($\sim6\%$ compared with $\sim22\%$, albeit with a higher scatter ($\sigma=0.12$ compared with $\sigma=0.077$), suggesting that while we are less accurate than others, we are having fewer estimates catastrophically fail. This optimisation of outlier rate (reducing the number of estimations that catastrophically fail) is motivated by the science goals of the \ac{EMU} Project. We note that this comparison should be further tempered by the fact that the differing training and test set sizes, and the redshift distribution of the overall datasets have a large impact on the overall statistics.

\begin{figure}
    \centering
    \includegraphics[trim=0 0 0 0, width=0.5\textwidth]{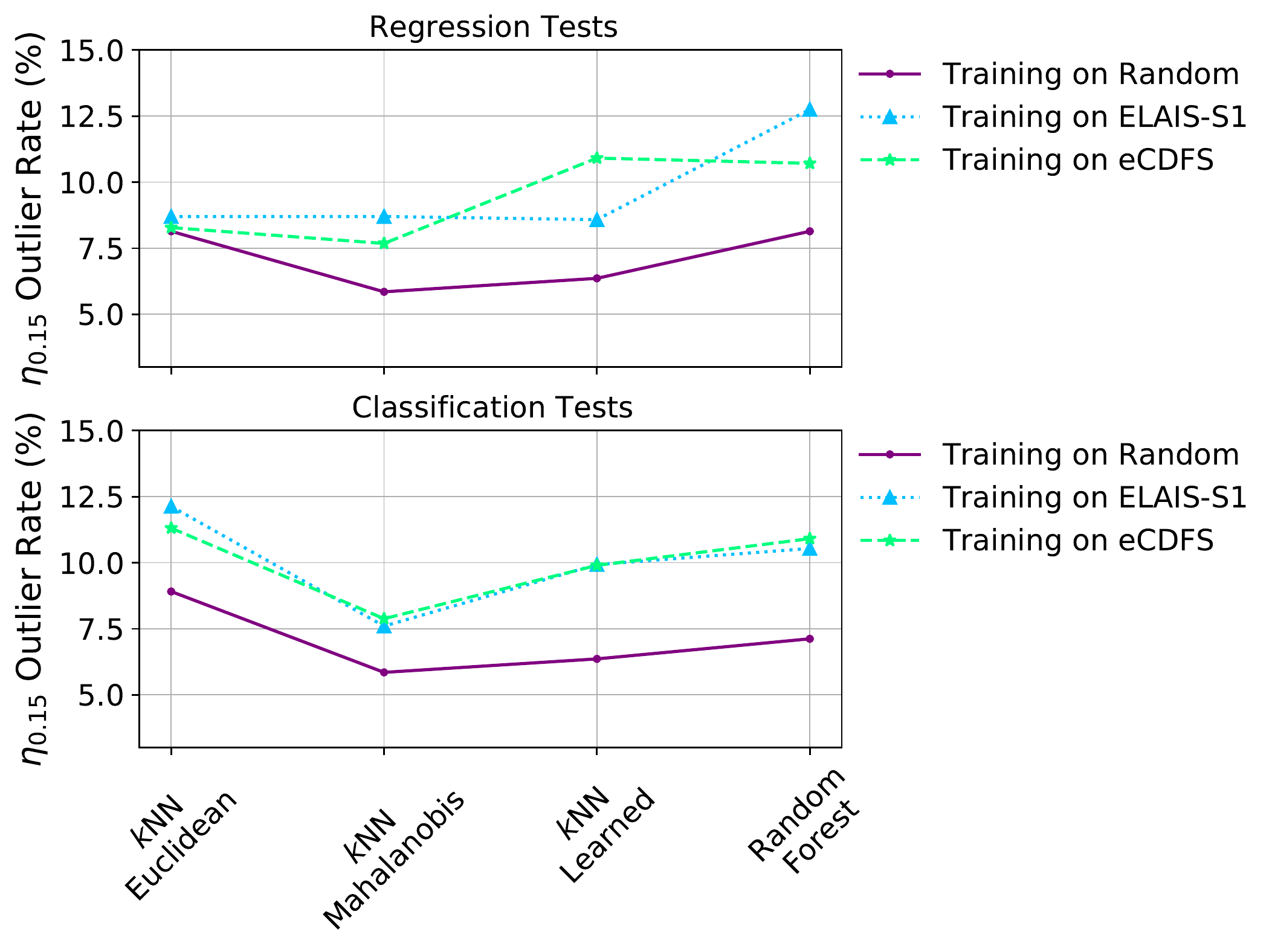}
    \caption{A comparison of the $\eta_{0.15}$ outlier rates presented in Section~\ref{sec:results}. The top panel compares the regression-based tests, with the bottom panel comparing the classification-based tests. The x-axis shows the different tests run, with the y-axis showing the $\eta_{0.15}$ outlier rate. The different colours/styled lines represent the different training data used -- the purple solid line used a training sample randomly selected from both the ELAIS-S1 and eCDFS fields, the blue dotted line used a training sample comprised of galaxies in the ELAIS-S1 field, and the the greed dashed line used a training sample comprised of galaxies in the eCDFS field. }
    \label{fig:comparison}
\end{figure}

\section{Conclusion}

We have used a radio-selected point-source catalogue consisting of ATLAS radio data, DES optical photometry, SWIRE infrared data and OzDES spectroscopic redshifts to test the kNN and RF algorithms for estimating the redshift of galaxies. 

We have used both regression and classification modes of these algorithms in order to balance highly unbalanced training sets, and have shown that using classification modes increases the effectiveness of the algorithms at a redshift of $z > 1$ (noting that while we are generally able to identify the galaxies at $z > 1$, we are unable to estimate their exact redshift). Given this increase in effectiveness at isolating the high-redshift galaxies using classification methods, a mix of classification- and regression-based methods would provide the best possible result over larger redshift ranges, allowing the higher redshift galaxies to be identified, while still being able to estimate the redshift of nearby galaxies to high accuracy. 

In our tests we have shown that when using the classification modes of the algorithms, the \ac{kNN} algorithm using the Mahalanobis distance metric performs better than the alternative methods tested. We particularly note that the $z > 1.02$ bin was correctly predicted in 74\% of cases -- far better than the regression regimes that fail consistently at that range.

When using the regression modes of the algorithms, the \ac{kNN} algorithm using the Mahalanobis distance metric performed statistically significantly better than most of the alternate methods tested (\ac{kNN} using Euclidean Distance: p value = 0.0183, and the \ac{RF} algorithm 0.0183) -- the \ac{kNN} algorithm paired with the Mahalanobis-like \ac{MLKR} distance metric was statistically insignificantly different (p = 0.5750). In both regression and classification methods, the \ac{kNN} algorithm outperforms the much more widely used \ac{RF} algorithm.

Finally, we tested whether there would be a significant difference in the $\eta_{0.15}$ outlier rate between a model trained and tested on randomly split data, and a model that is trained on one field of sky and tested on another. While the results were generally suggestive of the random training sample out-performing the field-specific training samples, the difference is not statistically significant (Regression p value = 0.2072 and classification p value = 0.4397). This suggests that for new fields being observed with similar strategies to previously observed fields, any differences in measured photometry should produce minimal effect on the estimated redshifts. It should be noted, however, that this does not take into consideration any differences in photometry caused by differences in telescopes. For example, if a galaxy was to have SkyMapper $g$, $r$, $i$ and $z$ photometry instead of \ac{DES} photometry. To determine the impact of changes of telescopes on the data would require further testing, and will be the subject of further work utilising Transfer Learning. 

\section{Implications for Future Radio Surveys}

Traditional photometric template fitting methods typically struggle to estimate the redshift of radio galaxies. In the near future, large area radio surveys like the \ac{EMU} survey are set to revolutionise the field of Radio Astronomy, with the number of known radio galaxies set to increase by an order of magnitude. This work shows that broadband photometry at similar wavelengths to those available in present and near-complete all-sky surveys will be enough to estimate acceptable redshifts for $\sim$\,95\% of radio sources with coverage over those bands. Further, while continuous redshift values are particularly difficult to estimate for radio sources at high-redshift, they can still be identified as high redshift with the majority of sources correctly placed in the highest redshift bin using classification modes. 

\section*{Acknowledgments}
The Australia Telescope Compact Array is part of the Australia Telescope National Facility which is funded by the Australian Government for operation as a National Facility managed by CSIRO. We acknowledge the Gomeroi people as the traditional owners of the Observatory site.

Based in part on data acquired at the Anglo-Australian Telescope. We acknowledge the traditional owners of the land on which the AAT stands, the Gamilaroi people, and pay our respects to elders past and present.

This project used public archival data from the Dark Energy Survey (DES). Funding for the DES Projects has been provided by the U.S. Department of Energy, the U.S. National Science Foundation, the Ministry of Science and Education of Spain, the Science and Technology FacilitiesCouncil of the United Kingdom, the Higher Education Funding Council for England, the National Center for Supercomputing Applications at the University of Illinois at Urbana-Champaign, the Kavli Institute of Cosmological Physics at the University of Chicago, the Center for Cosmology and Astro-Particle Physics at the Ohio State University, the Mitchell Institute for Fundamental Physics and Astronomy at Texas A\&M University, Financiadora de Estudos e Projetos, Funda{\c c}{\~a}o Carlos Chagas Filho de Amparo {\`a} Pesquisa do Estado do Rio de Janeiro, Conselho Nacional de Desenvolvimento Cient{\'i}fico e Tecnol{\'o}gico and the Minist{\'e}rio da Ci{\^e}ncia, Tecnologia e Inova{\c c}{\~a}o, the Deutsche Forschungsgemeinschaft, and the Collaborating Institutions in the Dark Energy Survey.

The Collaborating Institutions are Argonne National Laboratory, the University of California at Santa Cruz, the University of Cambridge, Centro de Investigaciones Energ{\'e}ticas, Medioambientales y Tecnol{\'o}gicas-Madrid, the University of Chicago, University College London, the DES-Brazil Consortium, the University of Edinburgh, the Eidgen{\"o}ssische Technische Hochschule (ETH) Z{\"u}rich,  Fermi National Accelerator Laboratory, the University of Illinois at Urbana-Champaign, the Institut de Ci{\`e}ncies de l'Espai (IEEC/CSIC), the Institut de F{\'i}sica d'Altes Energies, Lawrence Berkeley National Laboratory, the Ludwig-Maximilians Universit{\"a}t M{\"u}nchen and the associated Excellence Cluster Universe, the University of Michigan, the National Optical Astronomy Observatory, the University of Nottingham, The Ohio State University, the OzDES Membership Consortium, the University of Pennsylvania, the University of Portsmouth, SLAC National Accelerator Laboratory, Stanford University, the University of Sussex, and Texas A\&M University.

Based in part on observations at Cerro Tololo Inter-American Observatory, National Optical Astronomy Observatory, which is operated by the Association of Universities for Research in Astronomy (AURA) under a cooperative agreement with the National Science Foundation.

This work is based on archival data obtained with the Spitzer Space Telescope, which was operated by the Jet Propulsion Laboratory, California Institute of Technology under a contract with NASA. Support for this work was provided by an award issued by JPL/Caltech

\bibliography{EstimatingRedshiftML}

\end{document}